\RequirePackage{ifpdf}

\documentclass[letter,11pt]{article}
\usepackage{jheppub}
\pdfoutput=1 

\usepackage{amsmath}
\usepackage{epsfig,multicol,bbm}
\usepackage{graphicx}
\usepackage{bm}
\usepackage{multirow}

\usepackage{subfigure}

\def\be{\begin{equation}}
\def\ee{\end{equation}}
\def\ba{\begin{equation}\begin{aligned}}
\def\ea{\end{aligned}\end{equation}}
\def\bfig{\begin{figure*}[htb]}
\def\efig{\end{figure*}}
\def\fig#1{Fig.\ \ref{#1}}

\def\eqn#1{(\ref{#1})}

\def\nb{{\bar{n}}}

\def\J{{\cal J}}
\def\B{{\cal B}}
\def\WB{{\overline W}}
\def\YB{{\overline Y}}

\def\nslash{\fmslash{n}}

\def\pslash{\fmslash{p}}

\def\qslash{\fmslash{q}}
\def\kslash{\fmslash{k}}
\def\epslash{\fmslash{\varepsilon}}

\def\ncol{$n$-collinear}
\def\nbcol{$\bar n$-collinear}

\def\nbar{\bar n}
\def\tr{{\text{Tr}}}

\def\OMIT#1{{}}
\def\sceti{SCET$_{\rm I}$}
\def\scetii{SCET$_{\rm II}$} 
\def\as4pi{{\alpha_s C_F\over 4 \pi}}
\def\bigO#1{{\mathcal{O}(#1)}}
\def\als{{\alpha_s}}
\def\oas{{\bigO{\als}}}

\def\pperp{{{\vec p}_\perp}}
\def\pn{P_n}
\def\pnb{P_\nb}
\def\Tr{{\rm Tr}\;}
\def\lqcd{\Lambda_{\rm QCD}}
\def\disc{{\rm Disc\;}}
\def\gE{{\gamma_E}}
\def\leff{{\cal L}_{\rm eff}}

\def\pperp{{{\vec p}_\perp}}

\def\bra#1{\left\langle #1 \right|}
\def\ket#1{\left| #1 \right\rangle}

\def\sp#1#2{#1\cdot#2}

\def\plusD#1{{\left[#1\right]_+}}
\def\rsnp#1#2{{\left[#1 #2\right]}}
\def\lsnp#1#2{{\left\langle#1 #2\right\rangle}}

\def\rlsnpA#1#2{{\left\langle#1\pm | #2\mp\right\rangle}}
\def\lrsnpA#1#2{{\left\langle#1\mp | #2\pm\right\rangle}}

\def\eperp{\xi}

\def\Jlo{{\cal J}^{(0)}}
\def\Jnlo#1{{\cal J}^{(1#1)}}
\def\etab{\bar\eta}
\def\tr{{\rm Tr}}

\makeatletter
\def\fmslash{\@ifnextchar[{\fmsl@sh}{\fmsl@sh[0mu]}}
\def\fmsl@sh[#1]#2{%
   \mathchoice
     {\@fmsl@sh\displaystyle{#1}{#2}}%
     {\@fmsl@sh\textstyle{#1}{#2}}%
     {\@fmsl@sh\scriptstyle{#1}{#2}}%
     {\@fmsl@sh\scriptscriptstyle{#1}{#2}}}
\def\@fmsl@sh#1#2#3{\m@th\ooalign{$\hfil#1\mkern#2/\hfil$\crcr$#1#3$}}

\makeatother

\def\ns{\fmslash{n}}
\def\nbs{\fmslash{\nbar}}

\title{Power Counting and Modes in SCET }
 
\author{Raymond Goerke}
\author{and Michael Luke}
\affiliation{Department of Physics, University of Toronto, Toronto, ON, Canada M5S1A7}
\emailAdd{rgoerke@physics.utoronto.ca}
\emailAdd{luke@physics.utoronto.ca}

\abstract{
We present a formulation of soft-collinear effective theory (SCET) in the two-jet sector as a theory of decoupled sectors of QCD coupled to Wilson lines. The formulation is manifestly boost-invariant, does not require the introduction of ultrasoft modes at the hard matching scale $Q$, and has manifest power counting in inverse powers of $Q$.  The spurious infrared divergences which arise in SCET when ultrasoft modes are not included in loops disappear when the overlap between the sectors is correctly subtracted, in a manner similar to the familiar zero-bin subtraction of SCET.   We illustrate this approach by analyzing deep inelastic scattering in the endpoint region in SCET and comment on other applications.}

\begin{document}

\maketitle



\section{Introduction}

Effective Field Theory (EFT) provides a powerful calculational tool to study multi-scale processes in quantum field theory. In an EFT, the theory is defined so that only the degrees of freedom with energies and momenta below some cutoff $\Lambda$ are included in the theory, while degrees of freedom above the cutoff are integrated out of the theory and their effects replaced with a series of operators in the effective Lagrangian. Familiar examples of this approach are four-fermi theory and Heavy Quark Effective Theory (HQET).

By construction, the infrared physics of the full theory is reproduced in a corresponding low-energy effective theory.  In its usual formulation, soft-collinear effective theory (SCET) \cite{Bauer:2000ew, Bauer:2000yr,
Bauer:2001ct, Bauer:2001yt, Bauer:2002nz, Beneke:2002ph,Beneke:2002ni}, an EFT appropriate for describing the dynamics of jets of particles with invariant mass much less than their energies, differs from the two previous examples because this feature is not manifest.  In addition to integrating out degrees of freedom above the cutoff of the EFT, the low-energy degrees of freedom are then split into various modes, which are differentiated by the scaling of the components of their four-momenta.\footnote{A similar splitting of the low energy degrees of freedom into modes occurs in non-relativistic QCD (NRQCD).}  In \sceti\ processes (for example, deep inelastic scattering \cite{Idilbi:2007ff,Becher:2006mr,Manohar:2003vb}, thrust \cite{Becher:2008cf,Hornig:2009vb,Abbate:2010xh,Abbate:2012jh}, and the endpoint of the photon spectrum in $B\to X_s\gamma$ \cite{Lee:2004ja,Beneke:2004in,Bosch:2004cb,Bauer:2001mh,Bauer:2000ew}) the low-energy degrees of freedom consist of collinear modes with large momenta directed along the direction of each hadronic jet and small invariant mass, and ultrasoft modes, whose momentum components scale isotropically and whose invariant mass is parametrically smaller than that of collinear modes.  Defining the usual light-like vectors $n=(1,0,0,1)$ and $\bar n=(1,0,0,-1)$ and working in light-cone coordinates
\begin{equation}
a^\mu={1\over 2} n\cdot a \; \nb^\mu+{1\over 2} \nb \cdot a\; n^\mu+a_\perp^\mu\equiv {1\over 2} a^+  \nb^\mu+{1\over 2} a^- n^\mu+a_\perp^\mu
\end{equation}
the momentum components of collinear modes in the $n$ direction are defined to scale as $p_n\sim(p_n^+, p_n^-, p_{n\perp})\sim Q(\lambda^2,1,\lambda)$, whereas the momenta of ultrasoft modes scale as $p_{\rm us}\sim Q(\lambda^2,\lambda^2,\lambda^2)$.  This defines the small parameter $\lambda$, and the effective theory is defined by expanding QCD amplitudes in powers of $\lambda$.
 Other processes --- \scetii\ observables ---\cite{Bauer:2002aj,Hornig:2009vb,Stewart:2010qs,Becher:2011pf,Tackmann:2012bt} 
 require soft modes, $p_s\sim Q(\lambda, \lambda, \lambda)$. More complex processes with additional scales  require additional modes \cite{Becher:2006mr,Bauer:2011uc,Procura:2014cba,Pietrulewicz:2016nwo}.  
By describing ultrasoft and collinear degrees of freedom by separate fields, soft-collinear factorization theorems which are valid in the collinear limit of QCD (see 
also \cite{Feige:2013zla,Feige:2014wja}) are manifest in SCET at leading order in $\lambda$.

Since ultrasoft modes just correspond to the infrared limit of collinear modes, by treating them as separate fields with distinct interactions, SCET does not manifestly treat infrared physics in the same way as full QCD.  Ultrasoft and collinear modes are instead handled very differently in the formalism:  ultrasoft quark fields are described by four-component QCD quark fields,  while collinear quark fields are
described by two-component spinors with complicated nonlocal interactions \cite{Bauer:2000yr}.  To avoid double counting the effects of ultrasoft modes must be subtracted from graphs containing collinear modes, a process known as ``zero-bin subtraction" \cite{Manohar:2006nz}. Furthermore, expanding QCD amplitudes in powers of $\lambda$ corresponds to expanding in the ratios of several scales simultaneously:  $\sqrt{p_n^2}/Q\sim\lambda$, $p_{us}^\mu/Q\sim\lambda^2$ and $p_{us}^\mu/\sqrt{p_n^2}\sim\lambda$.   This makes SCET quite different from more familiar EFT's such as four-fermi theory and HQET, where multi-scale problems are handled by a sequence of EFT's:  at the highest scale $Q$, fields with invariant masses above $Q$ are integrated out of the theory and amplitudes are expanded in powers of $\Lambda_i/Q$ where the $\Lambda_i$'s represent infrared scales which are parametrically smaller than $Q$.   The theory is then run down to the next scale (in this case $\lambda Q$), at which point particles with invariant mass above the cutoff are again integrated out of the theory, amplitudes are expanded in powers of $\Lambda^\prime_i/(\lambda Q)$, a new EFT is matched onto, and so on.  No  subdivision of the low-energy degrees of freedom in the effective theory into separate modes is required at any point.  

Subdividing infrared degrees of freedom into modes is also a frame-dependent procedure, since the momentum scaling for collinear and ultrasoft modes only holds in certain reference frames.  In general, the mode decomposition introduces a privileged frame (typically the centre of mass frame) which may have no physical significance in the problem, and breaks manifest Lorentz invariance.  While physical results should be independent of this choice -- for example, it was shown in \cite{Manohar:2003vb} that deep inelastic scattering (DIS) could be analyzed in SCET in either the target rest frame, where only ultrasoft and \ncol\ modes were required, or the Breit frame, where \ncol, \nbcol\ and ultrasoft modes were required -- the degrees of freedom may differ in different reference frames, so the theory is not manifestly frame independent.  Finally, since there are an infinite number of ways that momenta can be defined to scale, it is not always clear what modes are required to describe a given process, and there have been disagreements in the literature over the counting of modes \cite{Manohar:2003vb,Becher:2006mr}.

In this paper we present a simple formulation of SCET\footnote{Despite the fact that it has no explicit ultrasoft degrees of freedom, we will continue to call the effective theory with distinct low invariant mass sectors SCET.  } which avoids the complications discussed above.
In a previous paper \cite{Freedman:2011kj}, it was shown that ultrasoft and collinear modes in SCET may each be described by separate copies of the full QCD Lagrangian, coupled through an external current which is expanded in powers of $\lambda$. Here we extend this formalism to demonstrate that ultrasoft modes do not need to be explicitly included as separate degrees of freedom; as expected from the above discussion, they are included in the infrared of collinear degrees of freedom.  The theory is therefore defined by different sectors which are differentiated by the fact that the invariant mass of each sector is small while the invariant mass of particles in different sectors is large, with no reference to the scaling of different momentum components.  We therefore do not have to introduce a small parameter $\lambda$ to define power counting, but instead power-counting is in powers of $k_i/Q$, where the $k_i$ are infrared scales in the EFT.  In addition, the formalism is manifestly invariant under boosts along the $n$ and $\nb$ directions.  This is in contrast to the usual formulation of SCET, where the momentum mode scalings are only valid in one particular reference frame.

The formalism presented here is a simple extension of that in \cite{Freedman:2011kj}.  The effective theory below the hard scale $Q$ consists of separate $n$ and $\nb$ sectors coupled via an external current.  Each sector is described by QCD, and interactions between the sectors are described by Wilson lines. The theory therefore looks like that presented in \cite{Freedman:2011kj}, but with no explicit ultrasoft modes.  Ultrasoft modes were originally introduced into SCET to allow the theory to eliminate ultraviolet divergences which were sensitive to the infrared regulator \cite{Bauer:2000ew}, so it might be expected that eliminating ultrasoft modes in SCET would leave loop graphs ill-defined.  However, we show that when the overlap between the $n$ and $\nb$ sectors is subtracted from loop graphs, analogously to the familiar zero-bin subtraction in SCET, the theory correctly reproduces the infrared of QCD, and explicitly including ultrasoft modes is not required. 

In this paper we restrict our attention to process with two sectors, such as DIS, although it may be generalized to processes with multiple sectors (such as three-jet events or $qq\to qqX$ scattering). 
In Section \ref{eftsec} we discuss the matching of the external current relevant for DIS near $x=1$ onto SCET at the scale $Q$.  We discuss tree-level matching at leading and subleading orders, and illustrate that the power counting is given by dimensional analysis.  We illustrate that the theory is manifestly boost-invariant.  In Section \ref{overlapsec}  we discuss the theory at the loop level, and show that SCET may be renormalized without including explicit ultrasoft modes, as long as the overlap between the two collinear modes of the theory is consistently removed, a procedure which we refer to as ``overlap subtraction".   Finally, in Section \ref{dissect}, we illustrate the necessity of overlap subtraction in the operator product expansion for DIS and match SCET onto the parton distribution function at one loop.    We present our conclusions in Section \ref{sec:conc}.

\section{The Effective Theory for DIS at Tree Level}\label{eftsec}

\begin{figure}[tbh]
   \centering
  \includegraphics[height=0.15\textheight]{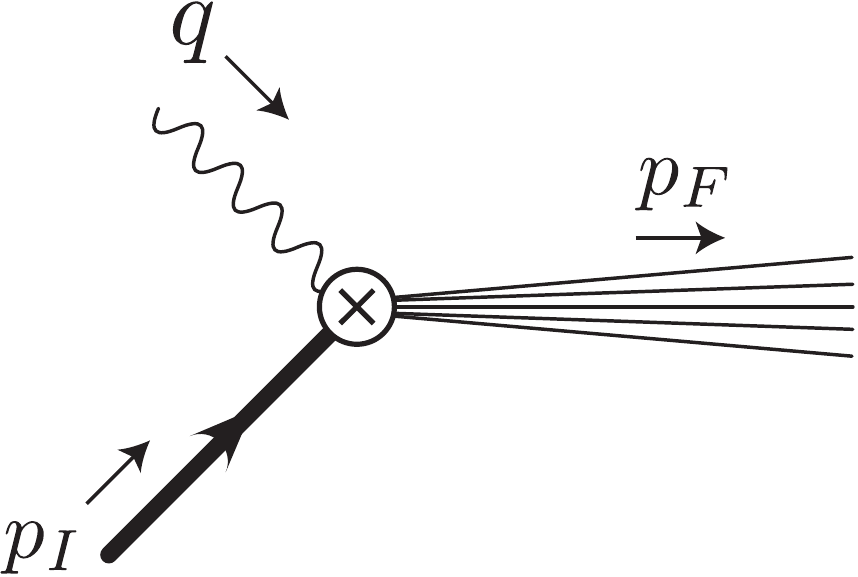}
   \caption{Deep inelastic scattering kinematics.}
   \label{disfig}
\end{figure}

For concreteness, consider DIS near the endpoint $x=1$, where 
\begin{equation} 
x\equiv -{q^2\over 2 P\cdot q},
\end{equation}
$q^\mu$ is the momentum transfer by the external current, and $P$ is the momentum of the incoming proton.  
The incoming state consists of low invariant mass partons with $p_I^2\sim\lqcd^2$, while the outgoing state consists of low invariant mass partons with $p_F^2\sim Q^2(1-x)\ll Q^2$, where $Q^2=-q^2\gg\lqcd^2$, as illustrated in \fig{disfig}.  There are thus two scales which are parametrically smaller than $Q$:  $\lqcd$ and $Q\sqrt{1-x}$.  DIS in this kinematic region
has been extensively studied in the framework of SCET, and provides an instructive example to illustrate our formalism.  

At renormalization scales $\mu<Q$, the renormalization scale of the theory is lower than the invariant mass $|p_I+p_F|$ of the total hadronic state, and QCD is matched onto SCET.  In our formalism, this theory consists of two distinct sectors coupled via an external current with an expansion in inverse powers of $Q$.  When the scale is run down below $\mu=Q\sqrt{1-x}$, the invariant mass of the final state is larger than the renormalization scale, and so the final state is integrated out of the theory by performing an operator product expansion (OPE) of the $T$-product of two currents onto a series of bilocal operators, consisting of the familiar parton distribution functions (PDF's) \cite{Collins:1981uw} and their higher-twist counterparts \cite{Ellis:1982cd}.  This sequence of EFT's is standard; our approach differs only in how the intermediate theory for $Q>\mu>Q\sqrt{1-x}$ is defined, so in the next few sections of the paper we will focus on this theory. 

As discussed previously, the standard formulation of SCET is frame-dependent, which leads to different descriptions in different reference frames.  In \cite{Manohar:2003vb}, DIS was analyzed in the framework of SCET in two different frames of reference, the target rest frame and the Breit frame.  In the target rest frame, the incoming partons were treated as ultrasoft and the outgoing partons as $n$-collinear, so only ultrasoft and $n$-collinear modes were needed in the calculation. In contrast, in the Breit frame the incoming partons were treated as $\nb$-collinear and the outgoing partons as $n$-collinear, so three modes, $n$-collinear, $\bar n$-collinear and ultrasoft, were required. The results in both frames were consistent, although the calculation also illustrated that the concept of scaling used was not entirely satisfactory, as there was no choice of $\lambda$ which corresponded to the DIS kinematics, since the invariant masses of the incoming and outgoing states are independent.  

In contrast, Ref. \cite{Becher:2006mr} argued that in order to treat the two infrared scales $\lqcd$ and $Q\sqrt{1-x}$ correctly, three distinct modes were required in both the Breit and target rest frames, with momenta scaling in the Breit frame as $p_{\bar c}\sim Q(1,\lambda^2,\lambda)$, $p_{hc}\sim Q(\epsilon, 1, \sqrt{\epsilon})$ and $p_{sc}\sim Q(\epsilon,\lambda^2,\sqrt\epsilon \lambda)$, where $\epsilon=1-x$ and $\lambda=\lqcd/Q$.  However, this description was also not entirely satisfactory, as graphs in which anti-collinear gluons were emitted by the incoming parton into the final state, and which by power counting produced large invariant mass final states, had to be excluded by hand. 

In this paper we argue that at a renormalization scale $\mu<Q$, this process may be most simply described in a boost-invariant manner by the effective Lagrangian
\begin{equation}\label{Leff}
\leff={\cal L}^n_{\rm QCD}+{\cal L}^\nb_{\rm QCD}+{\cal J}
\end{equation}
where ${\cal L}^n_{\rm QCD}$ and ${\cal L}^\nb_{\rm QCD}$ are two copies of the QCD Lagrangian with quark fields $\psi_{n,\bar n}$ and gluon fields $A^\mu_{n,\bar n}$.   
The SCET expansion arises through the interactions between the sectors, which are mediated by an external current with a large $O(Q)$ momentum transfer, 
 \begin{equation}\label{scetcurrent}
 {\cal J}=\Jlo+{1\over Q}{\cal J}^{(1)}+\dots \end{equation}
This differs from other formulations of SCET in several ways:
\begin{itemize}
\item Rather than being defined by scaling, the sectors are defined by a cutoff in their invariant masses:  $p_n^2\ll Q^2$, $p_\nb^2\ll Q^2$, while $p_n\cdot p_\nb\sim Q^2$, where $p_n$ and $p_\nb$ are the total momenta in the $n$ and $\nb$ sectors\footnote{We distinguish between $p_{n,\nb}$ and the momenta in the initial and final states $p_I$ and $p_F$  because, as will be discussed, partons in either sector may be present in both the initial and final states.}.   
Since the concept of collinear modes is frame-dependent -- the degrees of freedom in one sector only have large energies in the frame in which the other sector is soft -- we will henceforth simply refer to these as the $n$ and $\nb$ sectors, to distinguish them from the familiar collinear modes in SCET.

\item 
As in 
\cite{Freedman:2011kj}, there is no collinear expansion within the $n$ and $\bar n$ sectors.  Partons interact with one another via QCD: there is no small expansion parameter in the EFT within a sector.  The only source of $1/Q$ corrections is in the interaction between the sectors, which is mediated by the external current ${\cal J}$.  

\item The current $\cal J$ has an expansion in powers of $1/Q$, not $\lambda$.  ${\cal J}$ is constructed out of separately $n$ and $\nb$ gauge-invariant quantities, and as usual in SCET each sector sees the other as a Wilson line (or higher dimensional operators constructed from Wilson lines) - fields in the two sectors do not directly couple in ${\cal J}$. 

\item Ultrasoft degrees of freedom are not explicitly included as different fields in \eqn{Leff} -- different modes below the cutoff in a given sector are not described by different fields.  
In SCET, the interactions between the ultrasoft and collinear modes were expanded in a multipole expansion in $p_{us}/\sqrt{p_{n,\nb}^2}$; here no such expansion is performed since it is an expansion in two infrared scales of the theory.  (In contrast with \cite{Freedman:2011kj}, here we only introduce separate sectors when the invariant mass of those two sectors is of order $Q$.)  

\item The theory is manifestly boost invariant: that is, it is invariant under the rescaling $n\to \alpha n$, $\bar n\to \alpha^{-1} \bar n$.  Thus, there is no distinction in the EFT description of the process between frames which differ by a boost along the $n$-axis, such as the Breit and target rest frames.  

\end{itemize}
Note that in contrast with collinear and ultrasoft modes, particles in one sector typically have energy of order $Q$ with respect to particles in the other sector, and therefore are above the cutoff of the EFT and hence are integrated out of the effective Lagrangian for the other sector.  Thus, describing the $n$ and $\nb$ sectors with different fields does not in general corresponding to double-counting infrared physics. However, there is an important caveat here:  if an individual parton has momentum $p$ such that both $p\cdot p_n\ll Q^2$ and $p\cdot p_\nb\ll Q^2$, there is an ambiguity in assigning it to a sector, which will lead to a double-counting which must be subtracted.  This will be discussed in the next section. 
In the remainder of this section, we illustrate the formalism by considering tree-level matching onto SCET up to subleading order in $1/Q$.  

To match onto SCET at $\mu=Q$ we calculate corresponding matrix elements in QCD and SCET.  It is  convenient to consider matrix elements of states of definite helicity; this allows us to use standard spinor-helicity techniques \cite{Dixon:1996wi} to evaluate the matrix elements.  In addition, matching directly onto helicity eigenstates illustrates that it is unnecessary to assign power counting in $\lambda$ to different components of polarization vectors as is usually done in SCET; power counting in $1/Q$ in this theory is purely dimensional analysis. 

Consider the on-shell matrix element in QCD of an external electromagnetic current $J^\mu=\bar \psi\gamma^\mu \psi$ with momentum transfer $q$ between an incoming quark with momentum $p_1$ and an outgoing quark with momentum $p_2$ with $\pm$ helicity,
\begin{equation}\label{qcdmatrix}
\begin{aligned}
{\cal M}_{q\pm}&\equiv  \langle p_2\pm | J^\mu |p_1\pm\rangle\\
\end{aligned}
\end{equation}
(since the quarks are massless, for a vector current the helicities of the incoming and outgoing quarks are the same).  We choose our coordinates such that $\vec q_\perp=0$.
This matrix element is straightforward to evaluate using standard spinor-helicity methods; the calculation is described in Appendix \ref{helicityappendix}.  We obtain
\begin{equation}\label{fullQCDlo}
{\cal M}_{q^\pm}
=\sqrt{2 p_1^- p_2^+} e^{\mp i\varphi_{p}}\eperp_{\mp}^{\mu}
+\sqrt{2p_1^+ p_2^-}e^{\pm i\varphi_{p}}\eperp_{\pm}^{\mu}
+\sqrt{p_1^+p_2^+}\;\nb^{\mu}+\sqrt{p_1^- p_2^-}\;n^{\mu}
\end{equation}
where 
\begin{equation}
\eperp^{\mu}_{\pm} = (0,1,\mp i,0)
\end{equation} 
are transverse basis vectors 
and $\varphi_p$ is the polar angle of $\vec p_\perp$ in the $xy$ plane, 
\begin{equation}
e^{\pm i\varphi_p}\equiv {p^1\pm i p^2\over\sqrt{(p^1)^2+(p^2)^2}}={p^1\pm i p^2\over|\vec p_\perp|}
\end{equation}
and we have used the fact that $\vec p_{1\perp}=\vec p_{2\perp}\equiv \vec p_{\perp}$.  
The coordinates $p_i^\pm$ are frame-dependent, so there is no obvious small parameter in which to expand \eqn{fullQCDlo} which is valid in all frames.  In the SCET literature, one chooses a frame (typically the centre of mass frame), assigns power-counting in $\lambda$ to momenta and expands amplitudes in powers of $\lambda$.  However, each term in the expression (\ref{fullQCDlo}) is boost-invariant along the $\vec n$ direction:  under a change of reference frame, $n^\mu\to \alpha n^\mu$, $\nb^\mu\to \alpha^{-1}\nb^\mu$, we have $p_i^\pm\to \alpha^{\pm 1} p_i^\pm$, while the transverse basis vectors $\xi_\pm^\mu$ are unchanged, so each term in ${\cal M}_{q^\pm}$ is invariant.  It is convenient, then, to make boost invariance explicit by defining a new set of four-vectors $\eta$ and $\bar\eta$ by rescaling $n$ and $\nb$:
\begin{equation}\label{etas}
\begin{aligned}
\eta^{\mu} &= \sqrt{-\frac{q\cdot \nb}{q\cdot n}}\; n^{\mu}\\
\end{aligned}
,\quad
\begin{aligned}
\bar\eta^{\mu} &= \sqrt{-\frac{q\cdot n}{q\cdot \nb}}\; \nb^{\mu}\\
\end{aligned}
\end{equation}
These vectors are manifestly boost-invariant, and so $\frac{\sp{p_X}{\eta}}{Q} = (1-x)/x \sim (1-x)$ whether $n$ and $\nb$ were constructed in the target rest frame or the Breit frame, or indeed any other frame related to these by boosts along the $\vec{n}$ axis. Similarly, $\frac{\sp{p}{\bar\eta}}{Q}\sim \frac{\lqcd^2}{Q^2}$, $\frac{\sp{p_X}{\bar\eta}}{Q}\sim \frac{\sp{p}{\eta}}{Q}\sim 1$, $q\cdot\eta=-Q$, and $q\cdot\etab=Q$ in any frame boosted along the $\vec n$ axis. We can therefore define SCET in a manifestly boost-invariant way by expanding in $\frac{\sp{p_i}{\eta}}{Q}$ and ${\vec p_{i\perp}\over Q}$ for particles in the $n$ sector and $\frac{\sp{p_i}{\bar\eta}}{Q}$ and ${\vec p_{i\perp}\over Q}$ for particles in the $\nb$ sector. 
\OMIT{We distinguish this procedure from standard SCET power-counting, since we do not need to define the relative scaling of collinear versus soft or ultrasoft momentum modes, nor assign power counting to the fields.}
We thus can write \eqn{fullQCDlo} in the boost-invariant form,
\begin{equation}\label{fullQCDlob}\begin{aligned}
{\cal M}_{q^\pm}
&=\sqrt{2 p_1\cdot\etab\; p_2\cdot\eta}\; e^{\mp i\varphi_{p}}\;\eperp_{\mp}^{\mu}
+\sqrt{2p_1\cdot\eta\; p_2\cdot\etab}\;e^{\pm i\varphi_{p}}\;\eperp_{\pm}^{\mu}
+\sqrt{p_1\cdot\eta\; p_2\cdot\eta}\;\etab^{\mu}\\
&+\sqrt{p_1\cdot\etab\; p_2\cdot\etab}\;\eta^{\mu}.
\end{aligned}\end{equation}

To match onto SCET, we must identify the states in the effective theory corresponding to the full theory states.  SCET power counting is defined such that the incoming quark is properly described by the $\nb$ sector and the outgoing quark by the $n$ sector, so we
can then expand \eqn{fullQCDlob} and \eqn{fullqcdonegluon} in powers of $p_\perp/Q$ and $p_i\cdot\eta/Q$ to obtain
\begin{equation}\begin{aligned}\label{mqexp}
{\cal M}_{q^\pm}
&=Q\left(\sqrt{2} e^{\pm i\varphi_{p}}\eperp_{\pm}^{\mu}-{\sqrt{2}\over Q} e^{\pm i\varphi_{p}} \xi_{\pm}^\alpha p_{\perp \alpha}\left(\etab^\mu+\eta^\mu\right)+\dots\right)\\
\end{aligned}\end{equation}
where the ellipses denote terms of higher order in $1/Q$. 

SCET is constructed to reproduce the QCD result by expanding the current in powers of $1/Q$ via \eqn{scetcurrent}.  Since the $n$ and $\nb$ sectors just correspond to the $n$ and $\nb$ collinear modes in Refs.\ \cite{Freedman:2011kj, Freedman:2014uta}, the current will take the same form as in those papers.
At leading order in $1/Q$, the SCET current is \cite{Freedman:2011kj}
\begin{equation}\label{oldway}
\Jlo=C_2(\mu) O_2^{\mu}(\mu)=C_2(\mu) \left[\bar\psi_n  \overline W_n\right] P_\nb \gamma^{\mu} P_\nb \left[W_\nb^\dagger\psi_\nb\right]
\end{equation}
where
\begin {equation}
P_n={\ns \nbs\over4},\ P_{\nb}={\nbs\ns\over 4}
\end{equation}
are projection operators, 
$\psi_{n,\nb}$ and $A^\mu_{n,\nb}$ are full QCD fields in the given sector, and we have defined the Wilson lines (using the convention in \cite{Feige:2013zla})
\begin{equation}\begin{aligned}\label{w-wilson1}
W_\nb^\dagger(x)=&P \exp \left(ig\int_0^{\infty} n \cdot A^a_\nb(x+n s)T^a e^{-\epsilon s} ds\right)\\
\end{aligned}\end{equation}
for outgoing Wilson lines in the $\nb$ sector, and
\begin{equation}\begin{aligned}\label{w-wilson2}
\overline W_n(x)=&P\exp\left(ig\int_{-\infty}^0 \nb \cdot A^a_n(x+\nb s)T^a e^{\epsilon s} ds\right)\\
\end{aligned}\end{equation}
for incoming Wilson lines in the $n$ sector.  (Note that the subscript on the $W$'s denotes the sector with which the Wilson line interacts, not the direction of the Wilson line.)
Note that the Wilson lines $W_{n,\bar n}$ are invariant under rescaling $n\to \alpha n$, $\bar n\to\alpha^{-1}\bar n$, so the effective theory is manifestly boost invariant.

Following \cite{Kolodrubetz:2016uim}, it is convenient to rewrite $O_2$ so that its helicity structure is manifest.  We define gauge invariant quark operators with definite helicity\footnote{For processes with additional sectors, quarks in a given sector will couple to multiple Wilson lines in different directions, so the corresponding operators will not simply be constructed out of the $\chi_{n,\nbar}$ fields.  This is in contrast with collinear modes in SCET, which couple to a single Wilson line \cite{Chiu:2009yx}.}
\begin{equation}\begin{aligned}
\chi_{n}^{\pm}(x) = \WB^{\dagger}_{n}(x)P_{\pm}P_{n}\psi_{n}(x)\\
\chi_{\nb}^{\pm}(x) = W^{\dagger}_{\nb}(x)P_{\pm}P_{\nb}\psi_{\nb}(x)
\end{aligned}\end{equation}
where $P_{\pm}=(1\pm\gamma_5)/2$ and $a=n$ or $\bar n$.  
Using the relation
\begin{equation}\label{gammadecomp}
\gamma^{\mu} = {\fmslash{\nb}\over 2}n^{\mu}+ {\fmslash{\nb}\over 2}n^{\mu}-\eperp_-^{\mu}\fmslash{\eperp}_{+}-\eperp_+^{\mu}\fmslash{\eperp}_{-},
\end{equation} 
\eqref{oldway} becomes
\begin{equation}\label{newway}
O_2^\mu=-\eperp^{\mu}_{+}J_{n\nb+}^{ii}-\eperp^{\mu}_{-}J_{n\nb-}^{ii}
\end{equation}
where
\begin{equation}
J_{n\nb\pm}^{ij} = \bar\chi_{n}^{i\pm}\fmslash{\eperp}_{\mp}\chi_{\nb}^{j\pm}
\end{equation}
and $i$ and $j$ are colour indices.
We will often abbreviate $J_{n\nb\pm} \equiv J_{n\nb\pm}^{ii}$. 
Using these fields will simplify the construction of operators subleading in the $1/Q$ expansion (see \cite{Kolodrubetz:2016uim} and \cite{GoerkeInglis:2017}); in addition, matrix elements of operators between massless fields take a compact form when their helicity is specified so this is a natural operator basis.

Taking the matrix elements of \eqref{newway}, we find
\begin{equation}\begin{aligned}\label{scetLO}
\langle p_2(n) \pm | O_2^\mu | p_1(\nb)\pm\rangle =
&\eperp^{\mu}_{\pm}\sqrt{2 \sp{p_1}{\eta}\;\sp{p_2}{\bar{\eta }}}\;e^{\pm i\varphi_{p_1}}=
\sqrt{2}Qe^{\pm i\varphi_{p_1}}\eperp^{\mu}_{\pm}\left(1+O\left({p_\perp^2\over Q^2}\right)\right),\\
\end{aligned}\end{equation}
where we have explicitly labelled the particles in SCET by their sector. Comparing with \eqn{mqexp}, this reproduces the leading term with $C_2(\mu) = 1 + \oas$.  The subleading term in \eqn{mqexp} is reproduced by the operator\footnote{$O_2^{(1\perp)}$ and $O_2^{(1a)}$ (below) are linear combinations of the subleading operators introduced in \cite{Freedman:2011kj, Freedman:2014uta}.}
\begin{equation}
\Jnlo{\perp} = \frac{1}{Q}C^{(1\perp)}(\mu) O_2^{(1\perp)}(\mu)
\end{equation}
where
\begin{equation}\begin{aligned}\label{Jperp}
O_2^{(1\perp)}&=-\partial_n^{\alpha}\left(\bar\chi_n\gamma^\perp_\alpha\frac{\fmslash{\nbar}}{2}\gamma^{\mu}\chi_{\nb}\right)+\partial_{\nb}^{\alpha}\left(\bar\chi_n\gamma^{\mu}\frac{\fmslash{n}}{2}\gamma_{\alpha}^{\perp}\chi_{\bar n}\right)\\
&=-\bar\eta^{\mu}\left(i\left(\xi_+\cdot\partial_n\right)J_{n\nb+}+i\left(\xi_-\cdot\partial_n\right) J_{n\nb-}\right)\\
&\qquad\qquad+\eta^{\mu}\left(i\left(\xi_+\cdot\partial_\nb\right) J_{n\nb+}+i\left(\xi_-\cdot\partial_\nb\right) J_{n\nb-}\right),
\end{aligned}\end{equation}
which we have written in the usual form with Dirac matrices on the first line and in the helicity basis on the second.   
The subscripts on the derivatives $\partial_i$  indicate that the derivative only acts on fields in the $i$-sector; for example,
\begin{equation}
\left(\eperp_\pm\cdot \partial_n\right)J_{n\nb}^{\pm} = (\eperp_{\pm}^\mu\partial_\mu\bar\chi_n^{\pm})\fmslash{\eperp}_{\mp}\chi_{\nb}^{\pm}.
\end{equation}
Since the $\chi_{n,\nb}$ fields are gauge singlets, the derivatives in \eqn{Jperp} are regular, not covariant, derivatives.  Taking matrix elements of \eqn{Jperp} gives
\begin{equation}
\langle p_2(n) \pm | O_2^{(1\perp)} | p_1(\nb)\pm\rangle =-\sqrt{2} e^{\pm i\varphi_{p}}\eperp_\pm^\alpha p_{\perp\alpha}\left(\etab^\mu+\eta^\mu\right) 
\end{equation}
which reproduces the second term in (\ref{mqexp}) if $C^{(1\perp)}(\mu)=1+O(\alpha_s)$.  
We note that since $O_2^{(1\perp)}$ may be absorbed into $O_2$ by a small perpendicular rotation of $n$ and $\nb$, the relation $C^{(1a)}(\mu)=C_2(\mu)$ is determined by reparameterization invariance and will be true to all orders in $\alpha_s$.

At $O(1/Q)$, the effective current also contains operators of the form
\begin{equation}
{\cal J}\sim {C(\mu)\over Q} \B_{i\pm} J_{n\nb\pm}
\end{equation}
where \cite{Kolodrubetz:2016uim,Moult:2015aoa,Feige:2017zci} 
\begin{equation}\begin{aligned}
\B_{n\pm}^{ij} = \eperp_{\mp}^{\mu}(\WB_{n}^{\dagger}iD_{\mu}\WB_{n})^{ij}\\
\B_{\nb\pm}^{ij} = \eperp_{\mp}^{\mu}(W_{\nb}^{\dagger}iD_{\mu}W_{\nb})^{ij}
\end{aligned}\end{equation}
is a gauge invariant gluon field with definite helicity, which only contribute to states with external gluons.  We note that the power counting of this operator is just determined by dimensional analysis.  Thus, to match at subleading order we also must consider matching states with external gluons.  We define the matrix element with a gluon of momentum $k$ and polarization $\pm^\prime$ in the final state
\begin{equation}
{\cal M}_{q\pm g{\pm^\prime}}\equiv  \langle p_2\pm ; k\pm^\prime| J^\mu |p_1\pm\rangle.
\end{equation}
To simplify the amplitude, we choose the incoming state to have $\vec p_{1\perp}=0$, so that the total perpendicular momentum in the final state vanishes, since we have already determined the coefficient of $O_2^{(1\perp)}$. As shown in Appendix \ref{helicityappendix}, the one-gluon result can then be written in a boost-invariant form as
\begin{equation}\begin{aligned}\label{fullqcdonegluon}
{\cal M}_{q^\pm g^\pm}&= -\sqrt{2}T^a g\frac{\sqrt{\sp{p_2}{\bar\eta}}}{\sqrt{\sp{p_1}{\eta}}}\Big(\bar{\eta }^{\mu}+\eta^{\mu}+\sqrt{2}e^{\mp i\varphi_k} \frac{\sqrt{\sp{p_2}{\eta}}}{\sqrt{\sp{p_2}{\bar{\eta }}}}\xi_{\mp}^{\mu}+\sqrt{2}e^{\pm i \varphi_k}\frac{\sqrt{\sp{p_2}{\bar{\eta }}}}{\sqrt{\sp{p_2}{\eta}}}\xi_{\pm}^{\mu}\Big)\\
{\cal M}_{g^{\pm}q^{\mp}}&=-2gT^a e^{\mp i \varphi_k}\frac{
   \sqrt{\sp{p_1}{\eta}}}{\sqrt{\sp{p_2}{\eta}}}\xi_{\pm}^{\mu}.
\end{aligned}\end{equation}

\begin{figure}[tbh] 
   \centering
  \includegraphics[width=0.6\textwidth]{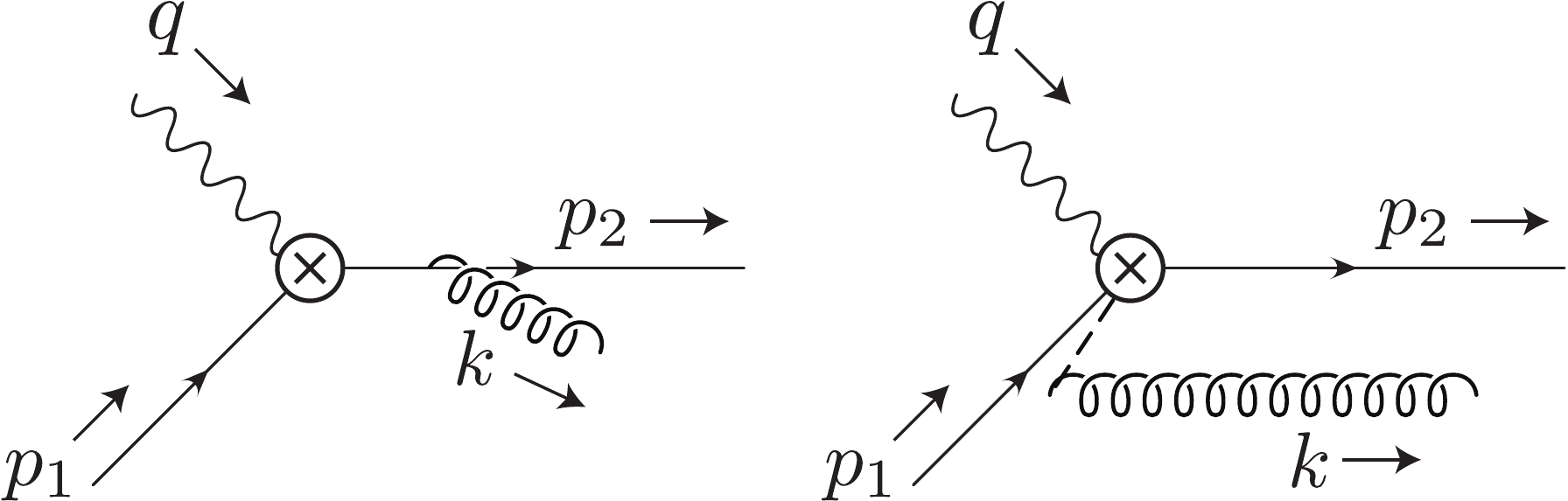}
   \caption{One-gluon matrix element of $O_2$.  The dashed line represents a Wilson line $\overline W_n$.}
   \label{o2onegluon}
\end{figure}

At leading order, the amplitude \eqn{fullqcdonegluon} is reproduced by the one-gluon matrix element of $O_2$, given by the diagrams in \fig{o2onegluon}, where the dashed line indicates an $n$-collinear gluon emitted from the nonlocal vertex of $O_2$.  This may be evaluated using the spinor-helicity formalism to give
\begin{equation}\label{o2onegluonmat}
\begin{aligned}
\langle p_2(n) \pm, k(n)\pm | O_2 | p_1(\nb)\pm\rangle &=-2g T^a {\sqrt{p_1\cdot\eta}\over Q}{p_2\cdot\etab\over\sqrt{p_2\cdot\eta}\sqrt{Q}} e^{i\varphi_k} \eperp^\mu_\pm \\
\langle p_2(n) \pm, k(n)\mp | O_2 | p_1(\nb)\pm\rangle &=-2gT^a e^{\mp i \varphi_k}\frac{
   \sqrt{\sp{p_1}{\eta}}}{\sqrt{\sp{p_2}{\eta}}}\xi_{\pm}^{\mu}.
   \end{aligned}
\end{equation}
Using the on-shell conditions $p_1\cdot\etab=0$, $(p_2+k)\cdot\etab=Q$ and $p_1\cdot\eta=Q+(p_2+k)\cdot\eta=Q+(p_2+k)^2/(p_2+k)\cdot\etab=Q+(p_2+k)^2/Q$, so to the order we are working we can take $p_1\cdot\eta=Q$.  The first matrix elements in \eqn{o2onegluonmat} therefore reproduce to relative order $1/Q$ the corresponding terms proportional to $\xi^\mu_\pm$ in \eqn{fullqcdonegluon}.

\begin{figure}[tbh] 
   \centering
  \includegraphics[width=0.6\textwidth]{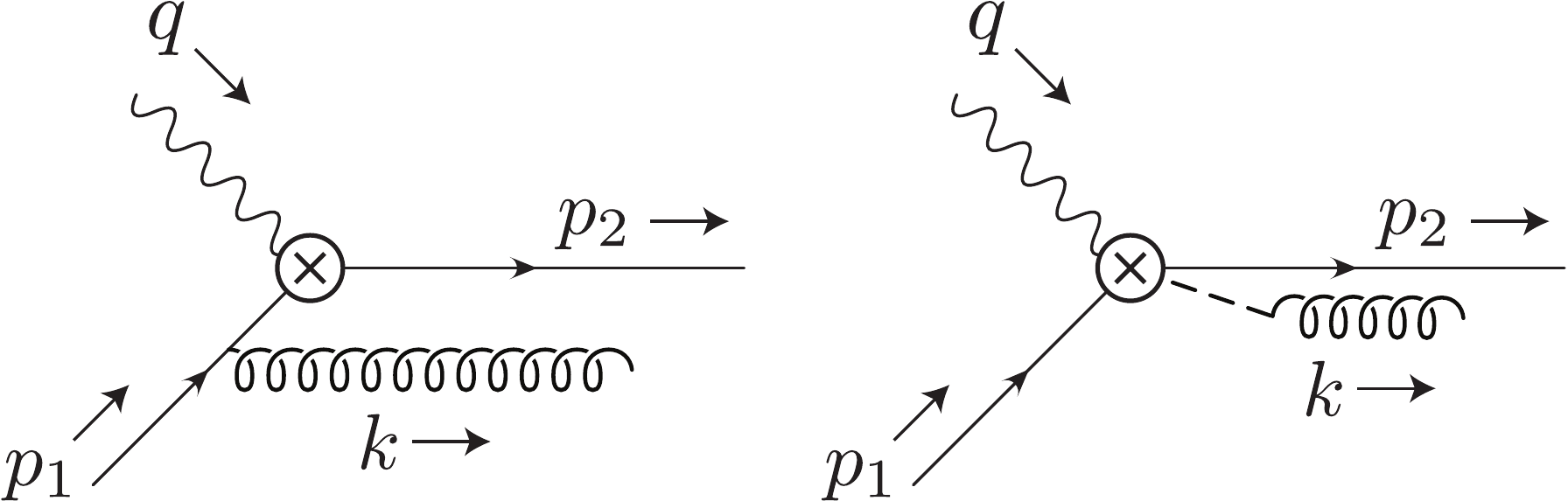}
   \caption{$\nb$ sector gluons emitted into the final state.  The dashed line indicates the Wilson line $W_\nb^\dagger$.}
   \label{wrongsector}
\end{figure}

Note that the QCD amplitude to produce a final state gluon has been reproduced by the amplitude in SCET to produce an $n$ sector gluon:  this is consistent with the power counting in the theory, where the invariant mass of the final state is much less than $Q$, and hence is described by the $n$ sector.  However, SCET also allows $\nb$ sector gluons to be emitted into the final state, through the diagrams in \fig{wrongsector}.  If the gluon is hard and $\nb$-collinear, the invariant mass of the final state will be large and the gluon will not contribute to the final state near $x=1$, but if the gluon momentum is soft, it can contribute to the final state amplitude.  Since the probability to produce an outgoing gluon has already been properly accounted for, this contribution is spurious and must be subtracted.  We will discuss this procedure in the next section.

Continuing to match the QCD amplitude to produce a final state gluon with the SCET amplitude to produce an $n$ sector gluon to $O(1/Q)$, we note
that is no contribution from $O_2^{(1\perp)}$ since we have chosen the total perpendicular momentum in each sector to vanish.  The $O(1/Q)$ terms in \eqn{mqexp} are reproduced by the subleading current  
\begin{equation}
\Jnlo{a}=\frac{1}{Q}C^{(1a)}(\mu) O^{(1a)}_2(\mu)
\end{equation}
where
\begin{equation}\begin{aligned}\label{Ja}
O^{(1a)}_2(\alpha) &= \bar{\mathcal{\chi}}_n(x) \B_{n}^{\alpha} \left(\gamma^{\mu}\frac{\fmslash{n}}{2}\gamma^\perp_\alpha-\gamma^\perp_\alpha\frac{\fmslash{\nbar}}{2}\gamma^{\mu}\right) \chi_\nb(x)\\
&= (\bar\eta^{\mu}+\eta^{\mu})\left(\B_{n-}^{ij}J_{n\nb+}^{ij}+\B_{n+}^{ij}J_{n\nb-}^{ij}\right).\\
\end{aligned}\end{equation}
The matrix elements of $\Jnlo{a}$  to produce an $n$--collinear gluon are easily calculated,
\begin{equation}\label{SCET_LO_matrixelement}
\begin{aligned}
\langle p_2(n) \pm, k(n)\pm | O_2^{(1a)} | p_1(\nb)\pm\rangle &=
-C_2^{(1a)}\sqrt{2} g \sqrt{\frac{\sp{p_2}{\bar{\eta }}}{Q}} (\eta^{\mu}+\bar{\eta }^{\mu})\\
\langle p_2(n) \pm, k(n)\mp | O_2^{(1a)} | p_1(\nb)\pm\rangle &= 0
\end{aligned}
\end{equation}
which gives $C_2^{(1a)} = 1 + \bigO{\als}$. 
Thus, the QCD amplitude to produce final state gluons is reproduced order by order in $1/Q$ by the effective theory amplitude to produce $n$ sector gluons.  Similarly, amplitudes with additional gluons in the initial state will be reproduced in SCET by amplitudes with $\nb$ sector gluons.  The leading amplitudes will be reproduced by the $\nb$ sector gluon matrix elements of $O_2$, while at $O(1/Q)$ SCET will also include contributions from operators with $\B_\nb$ fields, whose coefficients may be determined by considering incoming states with $\nb$ sector gluons.  The complete set of operators to $O(1/Q^2)$ relevant for dijet production in SCET is renormalized in \cite{GoerkeInglis:2017}.

\section{One Loop Matching and Overlap Subtraction}\label{overlapsec}

At the loop level, na\"\i vely applying the Feynman rules of the previous section leads to difficulties with the effective theory.  Matching onto $O_2$ at one loop is equivalent to matching onto standard SCET with only $n$ and $\nb$ collinear modes but no ultrasoft modes, which is known to be inconsistent: ultrasoft modes were originally introduced in SCET in \cite{Bauer:2000ew} in order to remove mixed ultraviolet-infrared divergences at one loop that cannot be absorbed in a process-independent counterterm.  Subsequently, it was shown in \cite{Manohar:2006nz} that to consistently reproduce the infrared physics of QCD, the contribution to SCET loop integrals from the region where collinear modes overlapped with ultrasoft modes must be subtracted, a procedure known as zero-bin subtraction.  Zero-bin subtracted graphs were then shown to be equivalent to subtracting the ultrasoft contribution to loop integrals \cite{Idilbi:2007ff,Idilbi:2007yi}. In \cite{Chiu:2009yx} it was shown that this procedure allowed SCET to be regulated in the infrared with a gluon mass.  Individual Feynman diagrams renormalizing the two-jet current with a massive gluon are not separately well-defined, but it was shown in \cite{Chiu:2009yx} that the sum 
\begin{equation}\label{zerobin}
R=I_n+I_\nb-I_0
\end{equation}
where $I_n$, $I_\nb$ and $I_0$ are the \ncol, \nbcol\ and  zero-bin subtracted diagrams (the latter quivalent to the ultrasoft contribution), is well-defined and reproduces the correct $m_g$ dependence in the amplitude.

We argue here that the formula \eqn{zerobin} arises naturally in our formulation of SCET, where now $I_n$ and $I_\nb$ are the one-loop diagrams renormalizing $O_2$ with $n$ and $\nb$ sector gluons, while the subtracted term $I_0$ subtracts the double-counting between the $n$ and $\nb$ sectors, rather than the zero-bin.  To distinguish this from zero-bin subtraction, we refer to this procedure as ``overlap subtraction."
\begin{figure}[tbh] 
   \centering
  \includegraphics[width=0.7\textwidth]{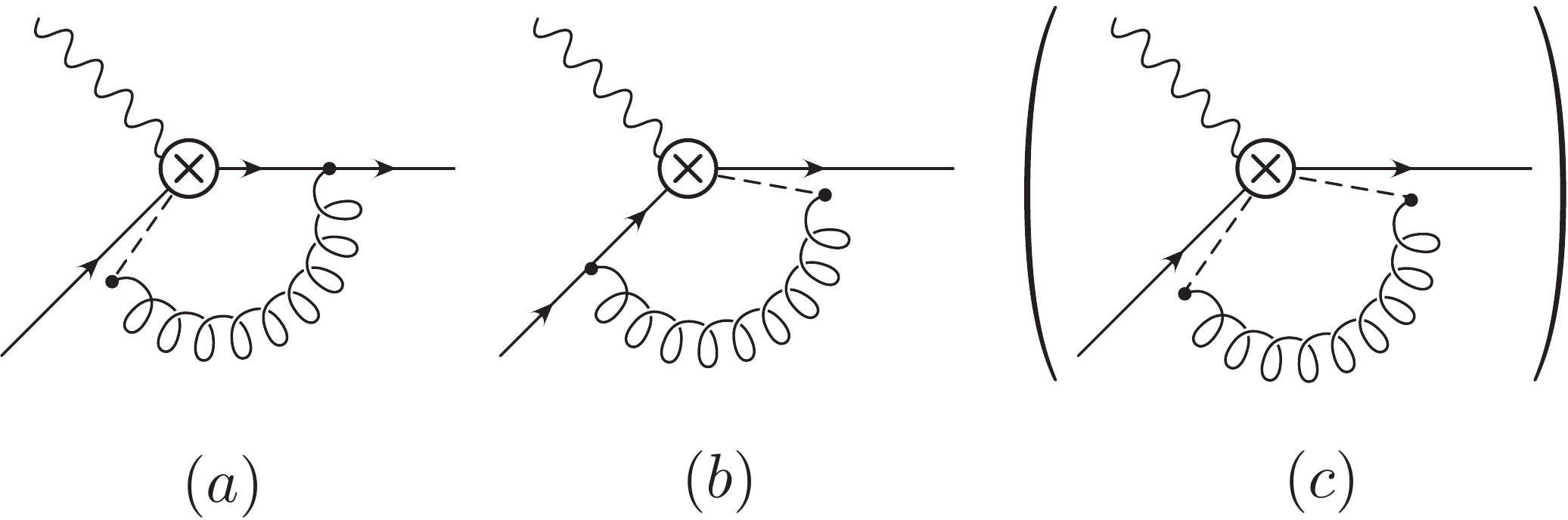}
   \caption{Renormalization of $O_2$.  Diagram (c) is the overlap subtraction. In (a), the gluon is in the $n$ sector; in (b), it is in the $\nb$ sector.  In (c), the dashed lines represent Wilson lines $Y_{n,\nb}$, depending on their direction. The divergences in (a) and (b) cannot be absorbed by a counterterm in SCET.  }
   \label{o2oneloop}
\end{figure}
In the one-loop renormalization of $O_2$, for example, both $n$ and $\nb$ sector gluons propagate through the corresponding loops in \fig{o2oneloop}.  This includes gluons with momentum $k$ for which both $k\cdot n\ll Q$ and $k\cdot \nb\ll Q$\footnote{Ultrasoft gluons, for example, meet these criteria.}, and so these gluons are included in both loop integrals, and therefore double-counted.  Subtracting the double-counting gives the formula (\ref{zerobin}).

This double-counting is already visible at tree level.  Cutting the graphs in \fig{o2oneloop} (a,b) gives graphs in which $n$ and $\nb$ sector gluons are emitted into the final state, respectively.  As discussed in Section \ref{eftsec},  the matrix element \eqn{o2onegluonmat} of $O_2$ with an $n$ sector gluon correctly reproduces the QCD amplitudes to emit a gluon into the final state at leading order.  The emission of an $\nb$ sector gluon in the final state, illustrated in \fig{wrongsector}, therefore double counts the degrees of freedom and must be subtracted from the EFT to reproduce QCD.   If the emitted gluon is hard and $\nb$-collinear, with $k\cdot\eta\sim O(Q)$, the invariant mass of the final state will be large the gluon won't contribute to the final state near $x\to 1$, but a gluon with momentum $k\cdot\eta\ll Q$ will be present in the final state in SCET.    Thus, without a subtraction, SCET will not correctly reproduce the infrared physics of QCD.

\begin{figure}[tbh]
\begin{center}
\includegraphics[width=0.6\textwidth]{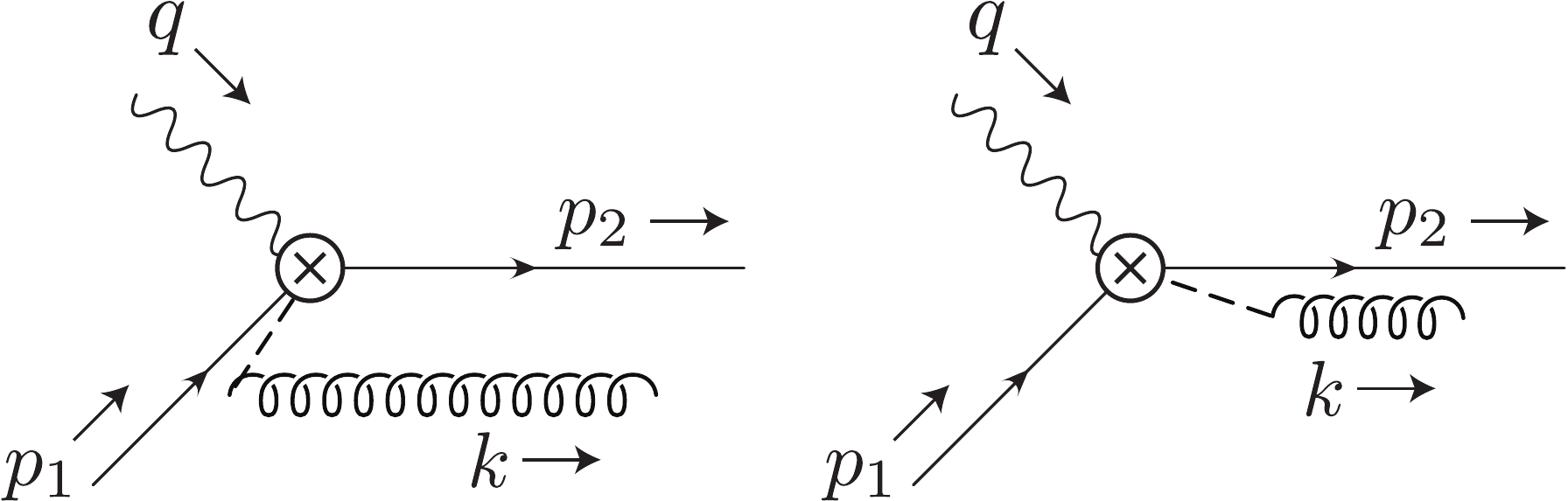}
\caption{Overlap subtraction graphs for producing a final-state gluon.}
\label{ovlp}
\end{center}
\end{figure}

The spurious amplitude to produce an $\nb$ sector gluon in the final state is easily calculated:
\begin{equation}\label{o2onegluonmatwrongpp}
\begin{aligned}
&\langle p_2(n) \pm, k(\nb)\pm | O_2 | p_1(\nb)\pm\rangle =\\
&\qquad -2g T^a \left(\sqrt{p_1\cdot\eta\over p_2\cdot\eta}e^{i\varphi_k} \eperp^\mu_++{p_2\cdot\etab\over (p_2-q)\cdot\etab}\sqrt{{p_2\cdot\eta\over p_1\cdot\eta}} e^{i\varphi_k}\eperp^\mu_++ \sqrt{p_2\cdot\eta\over p_1\cdot\eta} e^{i\varphi_k} \eperp^\mu_-\right) 
\end{aligned}\end{equation}
and
\begin{equation}\label{o2onegluonmatwrongpm}
\langle p_2(n) \pm, k(\nb)\mp | O_2 | p_1(\nb)\pm\rangle =-2gT^a \sqrt{p_1\cdot\eta\over p_2\cdot\eta}e^{-i\varphi_k}\eperp^\mu_+.
\end{equation}
Since the gluon is in the final state, SCET is constructed to reproduce QCD expanded in powers of $k_\perp/Q$ and $k\cdot\eta/Q$, so the second and third terms in \eqn{o2onegluonmatwrongpp} are subleading and may be neglected at leading order.  
The leading order amplitude is precisely that given by the diagrams in \fig{ovlp}, where the dashed lines are light-like Wilson lines
\begin{equation}\begin{aligned}\label{y-wilson}
Y^{\dagger}_n(x)=&P\exp\left(ig\int_0^\infty n \cdot A^a(x+n s)T^a ds\right)\\
\YB_\nb(x)=&P\exp\left(ig\int^0_{-\infty} \nb \cdot A^a(x+\nb s)T^a ds\right).\\
\end{aligned}\end{equation}
The notation in \eqref{y-wilson} is defined in analogy with ultrasoft Wilson lines in standard SCET, and so the subscript indicates the direction of the Wilson line (which is unfortunately opposite to the convention in \eqref{w-wilson1} and \eqref{w-wilson2}).   Unlike the $W_{n,\nb}$ Wilson lines, the gluon fields in $Y_n$ and $Y_\nb$ do not carry sector labels.  The gluons in \eqref{y-wilson} and \fig{ovlp} are distinct final states from the gluons in \fig{o2onegluon} and \fig{wrongsector}, and so the corresponding amplitudes do not interfere; instead, the contribution from the overlap graphs must be subtracted at the probability level.  Thus, the overlap gluons are similar to ghost fields in gauge theories.  As we have presented it at this stage, overlap subtraction is simply a prescription for avoiding double-counting, which doesn't come from the effective Lagrangian itself.  Schematically, for any sum over states, we make the subtraction
\begin{equation}\label{subtraction}
\int I_n d^4 k_n+\int I_\nb d^4 k_\nb\to \int I_n d^4 k_n+\int I_\nb d^4 k_\nb-\int I_0 d^4k
\end{equation}
where the integrand $I_0$ is obtained by taking the $n$ limit of the $\nb$ sector, or equivalently, the $\nb$ limit of the $n$ sector.   At tree level, where the sum over states is a phase space integral, this corresponds to subtracting the rate for emission of gluons from the overlap sector from the na\"\i ve SCET result, and so cancelling the effects of  $\nb$ sector gluon emission into the final state.

At one loop, the gluons in \fig{ovlp} are joined into a loop, which gives the contributions of double-counted gluons to the one-loop renormalization of $O_2$ to leading order.  The sum over states in \eqref{subtraction} is then a loop integreal, which leads to the prescription to subtract the diagram in \fig{o2oneloop}(c) from the one-loop renormalization, which is identical to the usual zero-bin subtraction prescription.
Regulating the effective theory with a gluon mass $m_g$, the one-loop renormalization of $O_2$ is then identical to that presented in \cite{Chiu:2009yx}.  As shown by those authors, the unsubtracted graphs $I_{n,\nb}$ with a massive gluon are not well defined in dimensional regularization, again indicating that the theory is not well-defined without overlap subtraction.  Including the overlap subtraction graph $I_0$, the sum (\ref{zerobin}) is well-defined, and gives
\begin{equation}\begin{aligned}
I_n+I_\nb-I_0-{1\over 2} w_n-{1\over 2}w_\nb=&{C_F\alpha_s\over 4\pi}\left[{2\over \epsilon^2}+{3-2\log{Q^2\over\mu^2}\over\epsilon}-\log^2{m_g^2\over\mu^2}\right.\\
&\hskip-1in\left.+2 \log{Q^2\over\mu^2}\log{m_g^2\over\mu^2}-3\log{m_g^2\over\mu^2}-{5\pi^2\over 6}+{9\over 2}\right]
\end{aligned}\end{equation}
where $I_{n,\nb}$ are the results of diagrams (a) and (b) in \fig{o2oneloop}, and the $w_{n,\nb}$'s are the wavefunction renormalization contributions from the $n$ and $\nb$ sectors.  This result
reproduces the nonanalytic dependence of the full QCD graph on $m_g$, and gives the counterterm
\begin{equation}\label{o2counter}
O_2^{c.t.}=-{C_F\alpha_s\over4\pi}\left[{2\over\epsilon^2}+{3-2\log{Q^2\over\mu^2}\over\epsilon}\right]
\end{equation}
which is independent of the infrared regulator and in agreement with the usual ultraviolet divergence in SCET.

It is also instructive to perform the calculation by including a $\Delta$ regulator \cite{Chiu:2009yx}, which corresponds to modifying the Wilson line Feynman rules in $W^\dagger_\nb$ and $\overline W_n$,
\begin{equation}
{1\over k\cdot n+i0^+}\to{1\over k\cdot n-\delta_n+i0^+},\ {1\over k\cdot \nb+i0^+}\to{1\over k\cdot \nb-\delta_\nb+i0^+}
\end{equation}
where $\delta_{n,\nb}$ are infrared regulators, which then makes the individual graphs well-defined.
The graphs in \fig{o2oneloop} (a,b) then give
\begin{equation}\begin{aligned}
I_{n}+I_{\nb}-{1\over 2} w_n-{1\over 2}w_\nb=& {C_F\alpha_s\over4\pi}\left[{1\over\epsilon}\left(3+2\log{\delta_n\delta_\nb\over Q^2}\right)+{1\over2}\log^2{\delta_n\over\delta_\nb}\right.\\
&\hskip-1in\left.-{1\over 2}\log{\delta_n\delta_\nb\over Q^2}\left(3\log{\delta_n\delta_\nb\over Q^2}+4\log{Q^2\over\mu^2}+3\right)-3\log{Q^2\over\mu^2}-{2\pi^2\over 3}+{7\over 2}\right].\\
\end{aligned}\end{equation}
 This result contains an ultraviolet divergence which depends on the infrared regulators $\delta_{n,\nb}$ and therefore cannot therefore be absorbed by a process-independent counterterm, illustrating that the EFT is not correctly reproducing the infrared behaviour of QCD.

Subtracting the overlap contribution $I_0$ in \fig{o2oneloop}(c) with the same regulators in the $Y_{n,\nb}$'s gives
\begin{equation}\begin{aligned}
I_n+I_\nb-I_0-{1\over 2} w_n-{1\over 2}w_\nb=& {C_F\alpha_s\over4\pi}\left[{2\over\epsilon^2}+{3-2\log{Q^2\over\mu^2}\over\epsilon} -{1\over 2}\log^2{Q^2\delta_n\delta_\nb\over\mu^4}+{1\over 2}\log^2{\delta_n\over\delta_\nb}\right.\\
&\hskip-1in\left.+{1\over 2}\left(4\log{Q^2\over\mu^2}-3\right)\log{Q^2\delta_n\delta_\nb\over\mu^4}-\log^2{Q^2\over\mu^2}-{\pi^2\over6}+{7\over 2}\right]  
\end{aligned}\end{equation}
in which the ultraviolet divergence is no longer infrared sensitive, and is cancelled by the standard SCET counterterm \eqref{o2counter}. This is, of course, exactly the same result as would be obtained in the usual formulation of SCET using the $\Delta$ regulator and a massless gluon.

As we have noted, the prescription \eqref{subtraction}, like the zero-bin subtraction, is a prescription for avoiding double-counting, not arising from the effective Lagrangian itself. However, it was shown in  \cite{Idilbi:2007ff,Idilbi:2007yi} that the zero-bin subtracted renormalization of $O_2$ is equivalent to dividing out the unsubtracted matrix element by a soft form factor given by the VEV of ultrasoft Wilson lines\footnote{This formula is also equivalent to the QCD factorization formula derived in \cite{Feige:2013zla,Feige:2014wja} after identifying soft and collinear Wilson lines and removing the soft sector from the current.}.  An analogous formula expresses the effect of overlap subtraction in our formalism:
\begin{equation}\label{factformula}
\langle p_2(n) | O_2|p_1(\nb)\rangle_{\rm subtracted} ={\langle p_2(n) | O_2|p_1(\nb)\rangle\over\frac{1}{N_c}\tr\langle 0|Y^\dagger_n \YB_{\nb}| 0 \rangle}.\end{equation}
Note that, formally, the numerator and denominator of \eqn{factformula} cannot be renormalized separately without including an additional infrared regulator such as the $\Delta$ regulator; only the sum of individual diagrams divided by the overlap term is well-defined in the effective theory. This definition of the subtraction also works for the renormalization of subleading dijet operators in SCET as well, as is demonstrated in \cite{GoerkeInglis:2017}, with the only modification being for certain operators the Wilson lines in the denominator must be in the adjoint rather than fundamental representation.

The overlap subtraction prescription is more complicated than \eqn{factformula} for $T$-products of currents (this is also true for zero-bin subtraction).  As we will show explicitly in the next section, diagrammatically performing the overlap subtraction for the $T$-product of two currents is necessary to correctly perform the phase-space integral in DIS, and is implemented by  
\begin{equation}\label{dissubtr}
\langle p_n|T O_2^\dagger(x)O_2^{\nu}(0)|p_n\rangle_{\rm subtracted}=\frac{\langle p_n|T O_2^\dagger(x)O_2(0)|p_n\rangle}{\frac{1}{N_c}\tr\langle 0| T \YB^{\dagger}_{\nb}(x)Y_{n}(x)Y^\dagger_{n}(0) \YB_{\nb}(0)|0\rangle}
\end{equation}
Comparing \eqref{dissubtr} to \eqref{factformula}, we see that each factor of $O_2$ in the numerator comes with a corresponding $Y_{n}^{\dagger}Y_{\nb}$ in the denominator. However, it is not sufficient to simply define an ``overlap subtracted'' current $O_2\to O_2/\langle 0|\frac{1}{N_c}Y^\dagger_n Y_{\nb}| 0 \rangle$ or overlap subtracted fields, since that does not correctly reproduce the sum over states in the denominator of (\ref{dissubtr}).  
For more complicated observables, such as event shapes which are defined by a measurement function, the overlap subtraction would have to be convoluted with the measurement function for the observable.

Finally, we note that this subtraction procedure only cancels the leading overlap in $1/Q$ overlap in \eqn{o2onegluonmatwrongpp}.  Cancelling the subleading terms will require including subleading operators in the overlap subtraction or, equivalently, in the denominators of Eqs.\ \eqref{factformula} and \eqref{dissubtr}.  This can be formally accomplished by matching tree-level probabilities with external gluons (note that probabilities, rather than amplitudes, must be matched since overlap graphs with external gluons have distinct final states in SCET from $n$ and $\nb$ gluons).  Ensuring that the tree-level probabilities are correctly reproduced in the EFT also provides a procedure to generalize overlap subtraction to situations with more than two sectors, such as three-jet events or $q\bar q \to q\bar q$ processes.

 \section{Matching onto the PDF}\label{dissect}

To obtain the familiar factorization theorem for DIS into hard, jet and parton distribution functions, we run the operators in SCET to the scale of the invariant mass of the final state, $\mu\sim Q\sqrt{1-x}$.  At this stage we perform an OPE to integrate out the $n$ sector and match onto an EFT containing only the $\nb$ sector (which is just QCD).  The corresponding bilocal operators in the EFT are the familiar parton distribution functions (PDF's) \cite{Collins:1981uw} and their higher-twist counterparts \cite{Ellis:1982cd}.  The matching at the scale $Q$ corresponds to the hard function, the matching condition at $Q\sqrt{1-x}$ the jet function and the PDF's are the soft function.  This is the procedure followed in \cite{Becher:2006mr,Manohar:2003vb}, but we will present it in some detail here because the SCET calculation differs somewhat in our formalism.

The hadronic tensor is defined as the totally inclusive DIS scattering cross section, and so by the optical theorem we can compute it by taking the discontinuity of the forward scattering amplitude. We therefore consider matrix elements of the discontinuity of the T-product of SCET currents,
\begin{equation}\begin{aligned}
T^{\mu\nu}&=\disc {1\over 2\pi} \int d^4 x\; e^{-i q\cdot x} T \J^\mu(x)^\dagger \J^\nu(0)\\
&=\disc {1\over 2\pi} \int d^4 x\; e^{-i q\cdot x} T \J^{(0)\mu}(x)^\dagger \J^{(0)\nu}(0)+O\left(1/Q\right)\\
&\equiv T^{(0)\mu\nu}+O\left(1/Q\right),
\end{aligned}
\end{equation}
where we have indicated that subleading currents, discussed in the previous sections, can be included to incorporate corrections due to the matching onto SCET at $Q$.  
At tree level, and leading order in the SCET expansion, we can compute the spin-averaged matrix elements between quark states, corresponding to the quark structure function:
\begin{equation}\label{disope1}
\begin{aligned}
{1\over 2}\sum_{\rm spins}\langle p|T^{(0)\mu\nu}|p\rangle&={1\over 2\pi}\disc{1\over 2}{\tr \pslash \pn \gamma^\nu\pn\left(\pslash+\qslash\right) \pnb\gamma^\mu\pnb\over (p+q)^2+i0^+}\\
&=-g_\perp^{\mu\nu}{1\over y}\left(1+{y\pperp^2\over Q^2}\right)\disc{1\over {1-y\over y}-{y\pperp^2\over Q^2}+i0^+}\\
\end{aligned}
\end{equation}
where
\begin{equation}
g_\perp^{\mu\nu}\equiv g^{\mu\nu}-{1\over 2}\left(n^\mu \nb^\nu+\nb^\mu n^\nu\right).
\end{equation}
and 
\begin{equation} y \equiv -q^+/p^+ \end{equation}
is the partonic equivalent of the $x$ variable in DIS.   
At the scale $Q^2(1-y)$ the invariant mass of the final state becomes larger than the invariant mass of incoming state, so perform an OPE by expanding the SCET result (\ref{disope1}) in powers of $\vec p_\perp^2/Q^2(1-y)$:
\begin{equation}\label{disope2}
\begin{aligned}
{1\over 2}\sum_{\rm spins}\langle p|T^{(0)\mu\nu}|p\rangle&=
&=-g_\perp^{\mu\nu}{1\over y}\left(1+{y\pperp^2\over Q^2}\right)\left[\delta(1-y)-{y^3 \vec p_\perp^2\over Q^2}\delta^\prime(1-y)+\dots\right]
\end{aligned}
\end{equation}
where the omitted terms are suppressed by more powers of $p_\perp/Q\sqrt{1-y}$, making explicit that this class of power corrections is proportional to inverse powers of the intermediate scale. 

The leading term in (\ref{disope2}) matches onto the bilocal quark distribution function
\begin{equation}\label{pdfmatching}
T^{(0)\mu\nu} \to -g_\perp^{\mu\nu}\int \frac{dw}{w}C_S(w)\phi(-q^+/w)+\dots
\end{equation}
where
\begin{equation}\begin{aligned}\label{loPDF}
\phi(r^+) &= \frac{1}{4\pi}\int_{-\infty}^{\infty}dt e^{-i  r^+ t}\bar\chi_{\nb}(nt)\fmslash{n}\chi_{\nb}(0)\\
&= \frac{1}{4\pi}\int_{-\infty}^{\infty}dt e^{-i  r^+ t}\bar\psi(nt)W(nt,0)\fmslash{n}\psi(0),
\end{aligned}\end{equation}
and in the second line we have dropped the $\nb$ subscripts since there is only one QCD sector in the theory.  We have also made explicit the fact that the outgoing and incoming semi-infinite Wilson lines partially overlap and leave a finite Wilson line, which gives the usual Collins-Soper definition of the PDF \cite{Collins:1981uw}. 

At tree-level, the spin-averaged matrix element of \eqref{loPDF} is
\begin{equation}
{1\over 2}\sum_{\rm spins}\langle p |\phi(-q^+/w)|p\rangle = \delta\left(1-\frac{y}{w}\right),
\end{equation}
which combined with the leading order term in \eqn{disope2} gives the matching condition
\begin{equation}\label{tree_CS}
C_S(w) = \delta(1-w) + \bigO{\alpha_s}.
\end{equation}

There are two sources of power corrections to \eqref{pdfmatching}. As already mentioned, we can add subleading currents $\Jnlo{a}$, etc. to the hadronic tensor $T^{\mu\nu}$. These are corrections at the hard scale matching $Q$, and will give corrections proportional to powers of $Q^2(1-y)/Q^2=1-y$ and $\lqcd/Q$. In addition, the OPE \eqref{disope1} can be carried out to higher orders, which will give corrections proportional to powers of $\lqcd/Q\sqrt{1-y}$.

\begin{figure}[tbh] 
   \centering
  \includegraphics[width=0.8\textwidth]{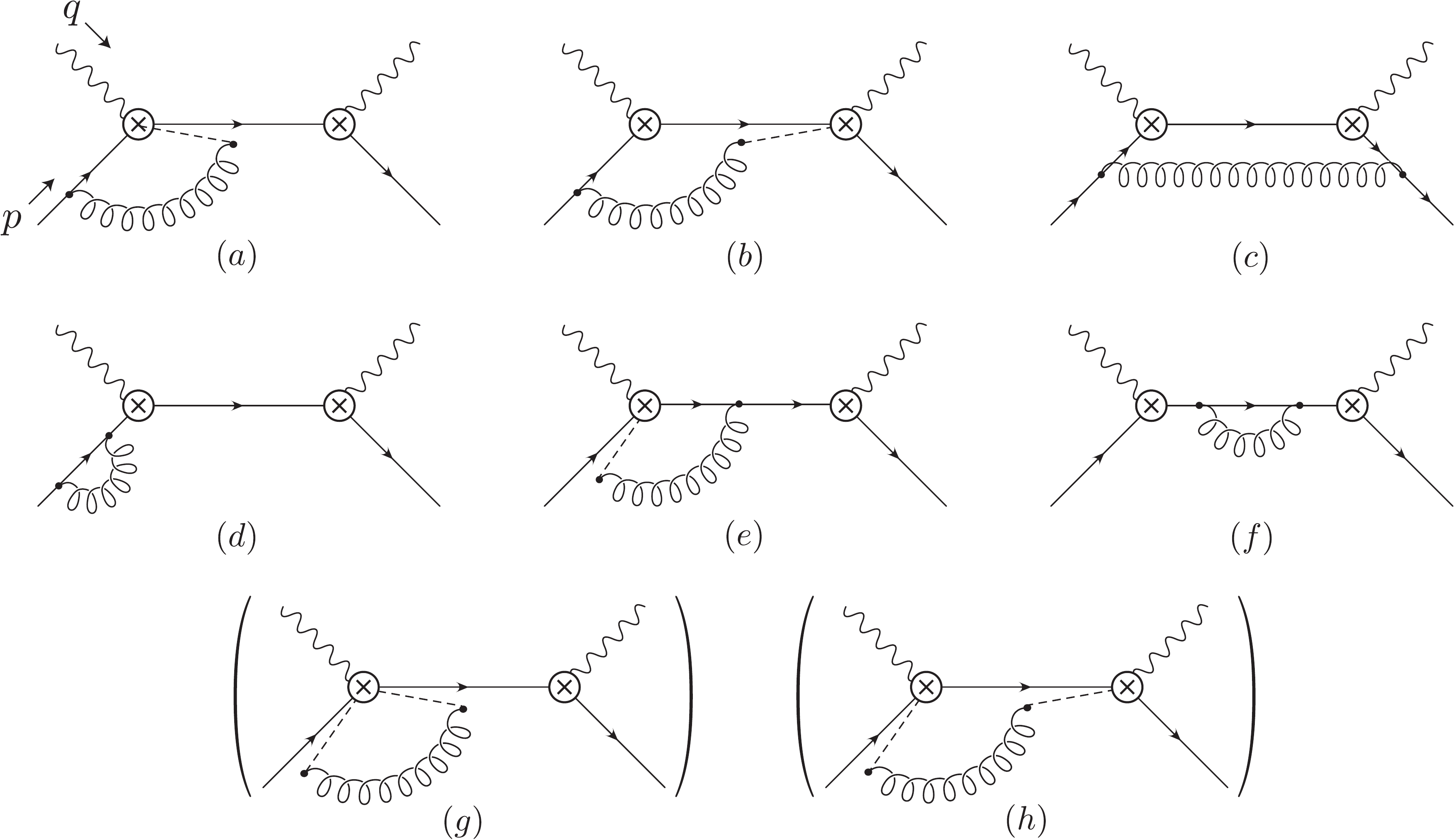}
   \caption{One loop corrections to the product of two electromagnetic currents in SCET.  Figures (g) and (h) are the overlap subtractions. The angle of the Wilson line indicates whether it is in the $n$ or $\nbar$ direction.}
   \label{disope}
\end{figure}

The rate for DIS at one loop in SCET is given by the diagrams in \fig{disope}.
Diagrams \fig{disope} (g) and (h) indicate the overlap subtraction graphs corresponding to the pairs \fig{disope} (a, f) and \fig{disope} (b,f).  As we will show below, the overlap graphs, along with their relative minus signs, are reproduced by the formula
\begin{equation}\label{dissubtr2}
\langle p_n|Tj^{\mu\dagger}(x)j^{\nu}(0)|p_n\rangle\rightarrow\frac{\langle p_n|T\J^{(0)\mu\dagger}(x)\J^{(0)\nu}|p_n\rangle}{\frac{1}{N_c}\tr\langle 0|T \YB^{\dagger}_{\nb}(x) Y_{n}(x)Y^\dagger_{n}(0) \YB_{\nb}(0)|0\rangle}.
\end{equation}
To see this, first note that the denominator, evaluated to one-loop, is
\begin{equation}\begin{aligned}
\frac{1}{N_c}\tr\langle 0|&T \YB^{\dagger}_{\nb}(x)Y_{n}(x)Y^\dagger_{n}(0) \YB_{\nb}(0)|0\rangle\\
=1&+2\int\frac{d^4k}{(2\pi)^4}\frac{2ig^2C_F}{\left(k^2+i\epsilon\right)\left(k^-+i\epsilon\right)\left(-k^++i\epsilon\right)}\\
&-e^{i k\cdot x}\int\frac{d^4k}{(2\pi)^4}\frac{2ig^2C_F}{\left(k^2+i\epsilon\right)\left(k^-+i\epsilon\right)\left(-k^++i\epsilon\right)}\\
&-e^{i k\cdot x}\int\frac{d^4k}{(2\pi)^4}\frac{2ig^2C_F}{\left(k^2+i\epsilon\right)\left(-k^-+i\epsilon\right)\left(k^++i\epsilon\right)}\\
&+\mathcal{O}(g^4),\\
\end{aligned}\end{equation}
while the numerator at tree-level is given by
\begin{equation}\begin{aligned}
\sum_{\text{spins}}\bra{p_1}T\J^{(0)\mu\dagger}(x)\J^{(0)\nu}(0)\ket{p_1}=-\int \frac{d^4p_2}{(2\pi)^4}\frac{4i p_2^-p_1^+ e^{-i(p_1-p_2)\cdot x}}{p_2^2+i\epsilon}g_{\perp}^{\mu\nu} + \mathcal{O}(g^2).
\end{aligned}\end{equation}
Combining these results and performing the integral over $x$, we find
\begin{equation}\begin{aligned}\label{ovlp_proof}
\int d^4x&e^{-i q\cdot x}\frac{\sum_{\text{spins}}\bra{p_1}T\J^{(0)\mu\dagger}(x)\J^{(0)\nu}(0)\ket{p_1}}{\frac{1}{N_c}\tr \bra{0}T \YB^{\dagger}_{\nb}(x)Y_{n}(x)Y^\dagger_{n}(0) \YB_{\nb}(0)\ket{0}}\\
\OMIT{=&-\int \frac{d^4p_2}{(2\pi)^4}\frac{4i p_2^-p_1^+ g_{\perp}^{\mu\nu}}{p_2^2+i\epsilon}(2\pi)^4\delta(q+p_1-p_2)\\
&+2\int \frac{d^4p_2}{(2\pi)^4}\frac{4i p_2^-p_1^+ g_{\perp}^{\mu\nu}}{p_2^2+i\epsilon}\int\frac{d^4k}{(2\pi)^4}\frac{2ig^2C_F}{\left(k^2+i\epsilon\right)\left(k^-+i\epsilon\right)\left(-k^++i\epsilon\right)}(2\pi)^4\delta(q+p_1-p_2)\\
&-\int \frac{d^4p_2}{(2\pi)^4}\frac{4i p_2^-p_1^+ g_{\perp}^{\mu\nu}}{p_2^2+i\epsilon}\int\frac{d^4k}{(2\pi)^4}\frac{2ig^2C_F}{\left(k^2+i\epsilon\right)\left(k^-+i\epsilon\right)\left(-k^++i\epsilon\right)}(2\pi)^4\delta(q+p_1-p_2-k)\\
&-\int \frac{d^4p_2}{(2\pi)^4}\frac{4i p_2^-p_1^+ g_{\perp}^{\mu\nu}}{p_2^2+i\epsilon}\int\frac{d^4k}{(2\pi)^4}\frac{2ig^2C_F}{\left(k^2+i\epsilon\right)\left(-k^-+i\epsilon\right)\left(k^++i\epsilon\right)}(2\pi)^4\delta(q+p_1-p_2-k)\\}
=&-\frac{4i (q^--p_1^-)p_1^+ }{(q-p_1)^2+i\epsilon}g_{\perp}^{\mu\nu}\\
&-2\frac{8g^2C_Fg_{\perp}^{\mu\nu} (q^--p_1^-)p_1^+ }{(q-p_1)^2+i\epsilon}\int\frac{d^4k}{(2\pi)^4}\frac{1}{\left(k^2+i\epsilon\right)\left(k^-+i\epsilon\right)\left(-k^++i\epsilon\right)}\\
&+8g^2C_Fg_{\perp}^{\mu\nu}\int\frac{d^4k}{(2\pi)^4}\frac{(q^--p_1^--k^-)p_1^+ }{(q-p_1-k)^2+i\epsilon}\frac{1}{\left(k^2+i\epsilon\right)\left(k^-+i\epsilon\right)\left(-k^++i\epsilon\right)}\\
&+8g^2C_Fg_{\perp}^{\mu\nu}\int\frac{d^4k}{(2\pi)^4}\frac{(q^--p_1^--k^-)p_1^+ }{(q-p_1-k)^2+i\epsilon}\frac{1}{\left(k^2+i\epsilon\right)\left(-k^-+i\epsilon\right)\left(k^++i\epsilon\right)}.\\
\end{aligned}\end{equation}
The first term on the RHS of (\ref{ovlp_proof}) is the amplitude for the tree level graph, the second line is the amplitude from \fig{disope}(g) and its mirror image, while the third and fourth lines correspond to the amplitudes for \fig{disope}(h) and its mirror.

We now procede to compute all of the graphs in \fig{disope}. \fig{disope}(a) is scaleless and vanishes in dimensional regularization, as does the wavefunction graph (d); \fig{disope}(b) gives for the spin-averaged matrix element
\begin{equation}\begin{aligned}
I_b=&\disc 2\alpha_s C_F f_\epsilon\int {d^dk\over(2\pi)^d} {1\over 2}{\Tr\pslash \pn\gamma^\nu\pn(\pslash-\kslash+\qslash)\pnb\gamma^\mu\pnb(\pslash-\kslash)\nslash\over ((p-k)^2+i0^+) ((p-k+q)^2+i0^+)(k^2+i0^+) n\cdot k}\\
=&-{\alpha_s C_F\over 4\pi}  g_\perp^{\mu\nu} \frac{\sqrt{\pi } 2^{2\epsilon-1 } e^{\gamma  \epsilon } \theta (1-y) Q^{-2 \epsilon } (1-y)^{-\epsilon -1} y^{\epsilon } (y (1-\epsilon )-\epsilon ) \Gamma (-\epsilon ) \mu^{2 \epsilon }}{\Gamma (1-\epsilon ) \Gamma \left(\frac{3}{2}-\epsilon \right)}\\
=&-{\alpha_s C_F\over 2\pi} g_\perp^{\mu\nu}\Bigg(\frac{\delta (1-y)}{\epsilon ^2}-\frac{1}{\epsilon}\left(\delta (1-y) \log \left(\frac{Q^2}{\mu^2}\right)+y \theta (1-y) \plusD{\frac{1}{1-y}}\right)\\
&+\frac{1}{4} \delta (1-y) \left(2 \log ^2\left(\frac{Q^2}{\mu^2}\right)-\pi ^2\right)+\theta (1-y) \plusD{\frac{1}{1-y}} \left(y \log \left(\frac{Q^2}{y \mu^2}\right)-y+1\right)\\
&+y \theta (1-y) \plusD{\frac{\log (1-y)}{1-y}}\Bigg)
\end{aligned}\end{equation}
where we have used the identity
\begin{equation}
{\theta(y-x)\over(y-x)^{1+\epsilon}}=-{1\over \epsilon}\delta(y-x)+
\theta(y-x)\left[{1\over
(y-x)_+}- \epsilon\left(\log(y-x)\over (y-x)\right)_++O(\epsilon^2)\right]
\end{equation}
to expand the results in $\epsilon$.
We have also used the relation
\begin{equation}\begin{aligned}&\int {d^{4-2\epsilon} k\over (2\pi)^{4-2\epsilon}}\delta(k^2)\delta\left((p+q-k)^2\right)
={2\theta(1-y)\over (4\pi)^{3-\epsilon}\Gamma\left(1-\epsilon\right)Q} \\
&\qquad \times \int_0^Q d k\cdot\nbar \int d k\cdot n\; (k\cdot\nbar)^{-\epsilon} (k\cdot n)^{-\epsilon}\delta\left(k\cdot n- {1-y\over y}(Q-k\cdot\nbar)\right).
\end{aligned}\end{equation}
to evaluate the imaginary parts of the various loop integrals.
Similarly, we find for \fig{disope}(c)
\begin{equation}\begin{aligned}
I_c=&\disc {g^2 f_\epsilon\over 2\pi}\int {d^dk\over(2\pi)^d} {1\over 2}{\Tr\pslash \gamma^\alpha(\pslash-\kslash)\pn\gamma^\nu\pn(\pslash-\kslash+\qslash)\pnb\gamma^\mu\pnb(\pslash-\kslash)\gamma_\alpha\over ((p-k)^2+i0^+)^2 (k^2+i0^+) ((p-k+q)^2+i0^+)}\\
=&{\alpha_s C_F\over 4\pi} T^{\mu\nu}\frac{(y-1) e^{\gamma  \epsilon } (\epsilon -1) \theta (1-y) Q^{-2 \epsilon } (1-y)^{-\epsilon } y^{\epsilon } \Gamma (3-\epsilon ) \Gamma (-\epsilon ) \mu^{2 \epsilon }}{\Gamma (3-2 \epsilon ) \Gamma (1-\epsilon )}\\
=&{\alpha_s C_F\over 4\pi} T^{\mu\nu}\Bigg(-\frac{(1-y) \theta (1-y)}{\epsilon }+\frac{1}{2} (1-y) \theta (1-y) \left(2 \log \left(\frac{Q^2}{y \mu^2}\right)+2 \log (1-y)-1\right)\Bigg)	
\end{aligned}\end{equation}

The $\nb$-collinear graphs, Figs. \ref{disope}(e) and (f) are also easily evaluated and give, respectively,
\begin{equation}\begin{aligned}
I_e=&{\alpha_s C_F\over 4\pi} T^{\mu\nu}\frac{e^{\gamma  \epsilon } \theta (1-y) Q^{-2 \epsilon } (1-y)^{-\epsilon -1} y^{\epsilon } \Gamma (2-\epsilon ) \Gamma (-\epsilon ) \mu^{2 \epsilon }}{\Gamma (2-2 \epsilon ) \Gamma (1-\epsilon )}\\
=&{\alpha_s C_F\over 2\pi}\Bigg(\frac{\delta (1-y)}{\epsilon ^2}+\frac{1}{\epsilon}\left(\delta (1-y) \left(1-\log \left(\frac{Q^2}{\mu^2}\right)\right)-\theta (1-y) \plusD{\frac{1}{1-y}}\right)\\
&+\frac{1}{4} \delta (1-y) \left(2 \log ^2\left(\frac{Q^2}{\mu^2}\right)-4 \log \left(\frac{Q^2}{\mu^2}\right)-\pi ^2+8\right)\\
&+\theta (1-y) \plusD{\frac{1}{1-y}} \left(\log \left(\frac{Q^2}{y \mu^2}\right)-1\right)+\theta (1-y) \plusD{\frac{\log (1-y)}{1-y}}\Bigg)\\
I_f=&{\alpha_s C_F\over 4\pi} T^{\mu\nu}\frac{\pi ^{3/2} 4^{\epsilon -1} e^{\gamma  \epsilon } \theta (1-y) Q^{-2 \epsilon } (1-y)^{-\epsilon -1} y^{\epsilon }(1-\epsilon) \csc (\pi  \epsilon ) \mu^{2 \epsilon }}{\Gamma (1-\epsilon ) \Gamma \left(\frac{3}{2}-\epsilon \right) \Gamma (\epsilon )}\\
=&{\alpha_s C_F\over 2\pi} T^{\mu\nu}\Bigg(-\frac{\delta (1-y)}{2 \epsilon }+\frac{1}{2} \delta (1-y) \left(\log \left(\frac{Q^2}{\mu^2}\right)-1\right)+\frac{1}{2} \theta (1-y) \plusD{\frac{1}{1-y}}\Bigg)
\end{aligned}\end{equation}
Note that there are also virtual contributions from \fig{disope}(e) (and its mirror) when the cut crosses right-hand quark propagator (left-hand in the mirror), and wavefunction contributions from \fig{disope}(f) when the cut crosses either of the outer quark propagators. Naively, these integrals are all sensitive to the scale $(p+q)^2\sim Q^2(1-y)$, but in each case the cut sets $(p+q)^2 = 0$, or $y=1$, and thus they are scaleless and vanish in dimensional regularization.

Finally, we compute the overlap subtractions, Figs.\ \ref{disope}(g) and (h).  \fig{disope}(g) is scaleless and vanishes, while (h) gives
\begin{equation}\begin{aligned}
I_h=&\disc {g^2 f_\epsilon\over 2\pi}\int {d^dk\over(2\pi)^d} {\Tr\pslash\pn\gamma^\nu\pn(\pslash+\qslash-\kslash)\pnb\gamma^\mu\pnb\over (k^2+i0^+) ((p-k+q)^2+i0^+)(-\nb\cdot k)(n\cdot k)}\\
=&{\alpha_s C_FT^{\mu\nu} e^\gE\over 4\pi}{y^\epsilon \Gamma(-\epsilon)\over(1-y)^{1+\epsilon}\Gamma(1-2\epsilon)}\left({\mu^2\over Q^2}\right)^\epsilon\\
=&{\alpha_s C_F \over 2\pi}T^{\mu\nu}\left({\delta(1-y)\over\epsilon^2}-{1\over \epsilon}\left(\log{Q^2\over\mu^2}\delta(1-y)+\left({1\over 1-y}\right)_+\right)\right)\\
&+\frac{1}{4} \delta (1-y) \left(2 \log ^2\left(\frac{Q^2}{\mu^2}\right)-\pi ^2\right)+\theta (1-y) \plusD{\frac{1}{1-y}} \log \left(\frac{Q^2}{y \mu^2}\right)\\
&+\theta (1-y) \plusD{\frac{\log (1-y)}{1-y}}
\end{aligned}\end{equation}

Combining these results with the relevant mirror image graphs gives the final result
\begin{equation}\begin{aligned}\label{finalresult}
I=&{\alpha_s C_F\over 2\pi}\left(\left({2\over\epsilon^2}+\frac{3-2\log{\mu^2\over Q^2}}{\epsilon}\right)\delta(1-y)-{1\over \epsilon}\left({1+y^2\over(1-y)_+}+{3\over2}\delta(1-y)\right)
\right.\\
&+\delta (1-y) \left(\log ^2\left(\frac{Q^2}{\mu^2}\right)-\frac{3}{2} \log \left(\frac{Q^2}{\mu^2}\right)-\frac{\pi ^2}{2}+7\right)-\frac{1}{2} (1-y) \theta (1-y)\\
&+\frac{1}{2} \theta (1-y) \plusD{\frac{1}{1-y}} \left(2 \left(y^2+1\right) \log \left(\frac{Q^2}{y \mu^2}\right)-(2 y^2+1)\right)\\
&+\left(y^2+1\right) \theta (1-y) \plusD{\frac{\log (1-y)}{1-y}}
\end{aligned}\end{equation}
The first two terms are cancelled by the ultraviolet counterterm in the effective theory, Eq.\ \eqn{o2counter}. The remaining divergences are infrared, and are related to the Altarelli-Parisi splitting kernel.

\begin{figure}[tbh] 
   \centering
  \includegraphics[width=0.6\textwidth]{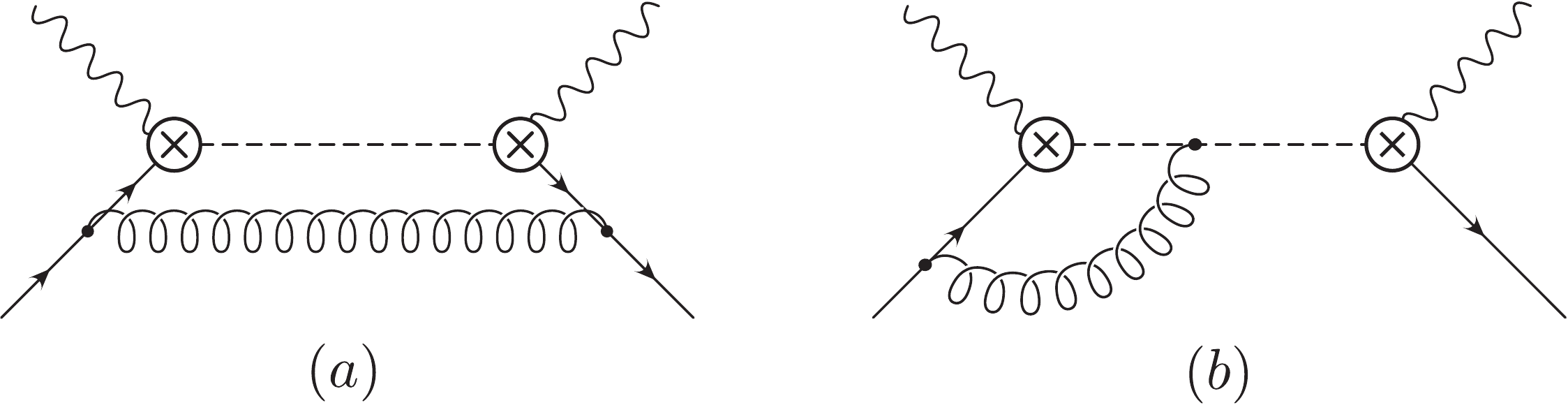}
   \caption{One loop corrections to the bilocal PDF distribution function.}
   \label{pdfgraphs}
\end{figure}

Having computed the structure functions at one loop in SCET, we compute the corresponding diagrams for the PDF so that we can determine $C_S$ in \eqref{pdfmatching} at one-loop. The relevant diagrams are shown in \fig{pdfgraphs}. These graphs are all scaleless and vanish in dimensional regularization. Thus, the matching condition $C_S$ at order $\alpha_s$ is given by \eqref{finalresult} after the ultraviolet divergences have been removed by the counterterms for the effective theory. As noted in the previous section, the first divergences in \eqref{finalresult} are removed by the counterterms for the currents appearing in $T^{\mu\nu}$; the remaining divergences, which are infrared in the effective theory above $Q^2(1-x)$, have been converted to ultraviolet divergences of the PDF and can be removed by an ultraviolet counterterm via the convolution
\begin{equation}
\phi^{(B)}(-q^+u) = \int\frac{dv}{v}Z\left(\frac{u}{v}\right)\phi^{(R)}(-q^+v)
\end{equation}
where
\begin{equation}
Z(z) = \delta(1-z) + \frac{\alpha_s}{4\pi}{1\over \epsilon}\left({1+z^2\over(1-z)_+}+{3\over2}\delta(1-z)\right),
\end{equation}
As is well-known, this counterterm corresponds to the standard Alterelli-Parisi splitting kernel.

$C_S$ can now be read off from the finite parts of \eqref{finalresult}:
\begin{equation}\begin{aligned}\label{cs_full}
C_S(w) &= \delta(1-w) +\frac{\alpha_sC_F}{2\pi}\Bigg(\delta (1-w) \left(\log ^2\left(\frac{Q^2}{\mu^2}\right)-\frac{3}{2} \log \left(\frac{Q^2}{\mu^2}\right)-\frac{\pi ^2}{2}+3\right)\\
&-\frac{1}{2} (1-w) \theta (1-w) +\frac{1}{2} \theta (1-w) \plusD{\frac{1}{1-w}} \left(2 \left(w^2+1\right) \log \left(\frac{Q^2}{w \mu^2}\right)-(2 w^2+1)\right)\\
&+\left(w^2+1\right) \theta (1-w) \plusD{\frac{\log (1-w)}{1-w}}\Bigg)
\end{aligned}\end{equation}

We can thus write down the full factorization formula for DIS in the endpoint region by including the matching conditions $C_2$ that relate the SCET currents $\Jlo$ to the QCD currents, i.e. for ${\cal J}^{\mu} = \bar\psi\gamma^{\mu}\psi$ and
 \begin{equation}
T^{\mu\nu} = \text{Disc}\frac{1}{2\pi}\int dx e^{-iq\cdot x}T{\cal J}^{\dagger\mu}(x) {\cal J}^{\nu}(0),
\end{equation}
we have
\begin{equation}
T^{\mu\nu} = g_{\perp}^{\mu\nu}H C_S\otimes\phi+\cdots,
\end{equation}
where $H = |C_2(\mu)|^2$ is the hard function, and we have used to shorthand $C_S\otimes \phi = \int dw/w \;C_S(w)\phi(-q^+/w)$. The ellipses denote subleading structure functions and subleading SCET operators.

It is useful to compare our calculation with other analyses of DIS in SCET. In \cite{Manohar:2003vb}, our calculation is most directly comparable to the calculation performed in the target rest frame. In that reference, it was argued that the graphs 7(b) and 7(c) (corresponding to \fig{disope}(a), (b) and (c) in this paper) had the same value as the corresponding graphs in the PDF calculation below $Q^2(1-x)$. This is because in that reference the propagator used for a collinear quark interacting with a ultrasoft gluon was the same as a gluon interacting with a Wilson line. In this work the quark propagator is a regular QCD propagator, and is therefore sensitive to the scale $(p+q)^2\sim Q^2(1-y)$, and so the contribution is not the same as the equivalent PDF graph \fig{pdfgraphs}(a).

This discrepancy is resolved by the overlap subtraction. We noted in section \ref{overlapsec} that the effect of the overlap subtraction is to remove the erroneous contributions of the incoming sector gluons being emitted into the final state. Indeed, graphs \fig{disope}(b) and (c) involve incoming sector gluons being emitted into the final state and so we expect this effect to be cancelled by the overlap subtraction. Although this cancellation is not immediately evident in the results above, we can consider the $y\rightarrow 1$ limit of graphs \fig{disope}(b) and (c):
\begin{equation}\begin{aligned}
\lim_{y\rightarrow 1}I_b+I_c=&\frac{\alpha_s C_F}{2\pi}T^{\mu\nu}\Bigg(\delta(1-y)\left(\frac{2}{\epsilon^2} - \frac{2\log\left(\frac{Q^2}{\mu^2}\right) + 1}{\epsilon} \log^2\left(\frac{Q^2}{\mu^2}\right) - \frac{\pi^2}{2}\right) \\
&+ 2\log\left(\frac{Q^2}{\mu^2}\right)\left[\frac{1}{1-y}\right]_+ + 2\left[\frac{\log(1-y)}{1-y}\right]_+\Bigg),
\end{aligned}\end{equation}
and compare them to the same limit of the overlap graph \fig{disope} (h),
\begin{equation}\begin{aligned}
\lim_{y\rightarrow 1}I_h=&\frac{\alpha_s C_F}{2\pi}T^{\mu\nu}\Bigg(\delta(1-y)\left(\frac{2}{\epsilon^2} - \frac{2\log\left(\frac{Q^2}{\mu^2}\right) + 1}{\epsilon} \log^2\left(\frac{Q^2}{\mu^2}\right) - \frac{\pi^2}{2}\right) \\
&+ 2\log\left(\frac{Q^2}{\mu^2}\right)\left[\frac{1}{1-y}\right]_+ + 2\left[\frac{\log(1-y)}{1-y}\right]_+\Bigg),
\end{aligned}\end{equation}
and so they do indeed cancel in the $y\rightarrow 1$ limit. Thus the net contribution from these graphs vanishes in this limit and is therefore equal to the contribution to the corresponding PDF graph \fig{pdfgraphs}(a), as was the case in \cite{Manohar:2003vb}. The fact that this cancellation is not exact is related to the fact that the overlap subtraction was only defined to remove the overlap at leading order in the SCET expansion. A subleading overlap prescription would presumably be able to reproduce the $\bigO{1-y}$ corrections to the cancellation.  The remaining graphs in Fig. 7 from \cite{Manohar:2003vb} correspond exactly with the corresponding graphs in \fig{disope} and the results are the same. Although we use different methods to regulate and extract the UV counterterm for the PDF, our final results also agree. 

In addition to the target rest frame, Ref.\ \cite{Manohar:2003vb} also performed the calculation in the Breit frame, where the incoming states were now treated as collinear. For the consistency of that formalism of SCET, an additional ultrasoft mode needed to be included in the Breit frame, and the author showed that the new effects of treating the incoming states as collinear were compensated by the effects of the additional ultrasoft sector, leading to the same results. We have showed in previous sections that this choice of reference frame has no effect on the sectors that must be included in the theory using our formalism, and  we have deliberately not specified which frame our calculations were performed in.

We can also compare to \cite{Idilbi:2007ff}, in which this calculation was performed by matching directly onto a factorization formula involving a soft function, jet function, and PDF. The jet function in that reference contains graphs equivalent to \fig{disope} (e) and (f), and since the remaining graphs in \fig{disope} sum to zero in the $y\rightarrow 1$ limit in our calculation, we expect that the matching condition $C_S(y)$ in \eqref{cs_full} should be equal to the jet function in that limit. Indeed, we find
\begin{equation}\begin{aligned}
\lim_{y \rightarrow 1} C_S(y) =& \delta(1-y) + \frac{\alpha_sC_F}{4\pi}\Bigg(\delta(1-y)\left(2\log^2\left(\frac{Q^2}{\mu^2}\right) - 3\log\left(\frac{Q^2}{\mu^2}\right)-\pi^2 + 7\right)\\
&+\left(4\log\left(\frac{Q^2}{\mu^2}\right) - 3\right)\left[\frac{1}{1-y}\right]_+ + 4\left[\frac{1}{1-y}\right]_+\Bigg),
\end{aligned}\end{equation}
which is exactly equal to the renormalized jet function in \cite{Idilbi:2007ff}. 

The soft function in \cite{Idilbi:2007ff} is similar to the object we defined in the overlap subtraction \eqref{dissubtr2}. A crucial difference is that in \cite{Idilbi:2007ff} the soft function is only convoluted with the external momentum along the light cone directions of the Wilson lines, whereas in our formula we convolute them in all directions. In that reference it was shown that the net effect of the soft function combined with the zero-bin subtractions in the jet function and PDF was to subtract the contribution from the soft function. In our formalism the overlap is subtracted directly, and while the contribution of the overlap is different form the soft contribution in \cite{Idilbi:2007ff}, our overlap contribution is cancelled in the $y\rightarrow 1$ limit by the graphs \fig{disope}(b, c), as discussed above, while the soft contribution in \cite{Idilbi:2007ff} is scaleless and vanishes, so our results are ultimately equivalent.

Finally, in \cite{Becher:2006mr}, the authors describe a two-step matching procedure that is very similar to the procedure we use here, where one first matches onto the SCET currents at the hard scale, and then matches onto the PDF at the intermediate scale, integrating out the final states and identifying the second matching coefficient with the jet function. These authors also identify two sources of power corrections coming from the insertion of subleading SCET currents and the inclusion of subleading parton distribution functions, agreeing with our analysis. The distinction between our approaches is in the identification of modes, as they describe new hard-collinear and soft-collinear modes in addition to the standard SCET anti-collinear modes which all must be included when matching onto SCET. The Feynman rules generated by the SCET Lagrangian thus constructed introduce diagrams in the one-loop computation of the hadronic tensor that must be removed by hand to maintain a consistent power-counting. These graphs are precisely the graphs \fig{disope}(b) and (c) in our formalism, and as shown above their effects are removed automatically at leading order in $(1-y)$ by the overlap subtraction.

\section{\label{sec:conc}Conclusions}

In this paper we have shown that SCET may be written as a theory of decoupled sectors, where the invariant mass of particles in each sector is much less than the hard scale $Q$, while the invariant mass of any pair of sectors is of order $Q$.  Fields in each sector are described by the QCD Lagrangian, and interactions are mediated by an external current, which has an expansion in powers of $\Lambda_i/Q$, where the $\Lambda_i$'s are the infrared scales in the theory.  We demonstrated that in order to consistently match probabilities from the full theory to SCET, the effects of overlapping degrees of freedom in each sector must be subtracted from loop and phase space integrals.  Once these effects are properly subtracted, the sum of loop graphs is well defined in the effective theory. 

In this work we restricted ourselves to the simplest situation in which there are only two sectors in the theory, and illustrated our results in the context of deep inelastic scattering; however, they will apply to any \sceti\  process - in particular, $B\to X_s\gamma$ and thrust distributions in the relevant kinematic regions.  In the former, SCET will take the form of a heavy (static) sector coupled to a final state $n$-sector through the external current, which matches at a lower scale onto light-cone distribution functions.  In the latter case, SCET will consist of $n$ and $\nb$ sectors in the final state, which is then matched onto the usual SCET soft function. The example of thrust is slightly complicated by the fact that the integration over final states is not inclusive, and so the usual cutting rules from the optical theorem must be modified, as described in \cite{Hornig:2009vb}. We expect that the overlap subtraction formula proposed in this paper will reproduce the correct results for thrust when this modified cutting prescription is used. 
It is also straightforward to extend this formalism to processes with multiple sectors, such as multi-jet events or $qq\to qqX$ scattering.  With multiple sectors, each sector will couple to multiple Wilson lines in different directions, similar to the way a soft field in the usual SCET formalism couples to multiple collinear fields.  The overlap subtraction procedure then  generalizes to eliminate double-counting between the various sectors.  Work on these examples is ongoing.

In future work, we will investigate the application of this formalism to \scetii\ observables, such as the massive Sudakov form factor and jet broadening.  We note in particular that events with small thrust ($\tau$) or jet broadening ($b_T$) are both dominated by dijet events with low invariant masses, and so SCET with $n$ and $\nb$ sectors has the correct degrees of freedom to describe the final states in both cases.  However, in standard SCET one must decide when matching at the scale $Q$ whether to match onto \sceti\ (collinear and ultrasoft modes) or \scetii\ (collinear and soft modes), whereas in our formalism the theories are identical at the scale $Q$.  The issue of summing the rapidity logs which arise in \scetii\ processes in this formalism is currently being investigated.

\begin{acknowledgments}
This work was supported by the Natural Sciences and Engineering Research Council
of Canada. We would like to thank C. Bauer, M. Inglis-Whalen, A. Manohar, J. Roy, M. Schwartz and A. Spourdalakis for discussions and comments on the paper.
\end{acknowledgments}

\OMIT{While the standard formulation of SCET has been shown to be consistent, it leads to several issues with the theory, both conceptual and practical.
From a conceptual point of view, in contrast to a cutoff, the ``scaling" of a mode is difficult to precisely define.  While the large component of a collinear momentum scales as $Q$, and so is of that order in the power counting, in loop and phase space integrals the components of momentum are integrated over all values. In particular, there are regions of integration where collinear and ultrasoft momenta overlap, which must be subtracted from the amplitude \cite{Manohar:2006nz}. 
Furthermore, the scaling of the invariant mass of a mode does not necessarily correspond to the renormalization scale of graphs containing that mode - for example, for the event shape variables described in [Lee et. al.]. SCET is used to factorize processes into various components - typically hard, jet and soft functions - each with its own characteristic renormalization scale. There is not always a simple relation between the characteristic scale of a jet function and the scaling of the collinear degrees of freedom - i.e. jet shapes.}

\OMIT{From a more practical point of view, in the standard formalism for an \sceti\ process, QCD is matched onto a theory of separate collinear and ultrasoft modes at the matching scale $Q$. However, there are two separate expansions going on at the matching scale: an collinear expansion in the small invariant mass of the final state jet over the hard scale, and an ultrasoft expansion in the ultrasoft momentum over the invariant mass of the jet. Both of these are expansions in powers of $\lambda$, but they are completely different expansions: the first is in power of $\sqrt{p_n^2}/Q$; the second in powers of $p_{us}^\mu/\sqrt{p_n^2}$. In the usual way of doing EFT's, these expansions would occur at the scales $\mu=Q$ and $\mu=\sqrt{p_n^2}$, respectively; however, in SCET they both occur at the matching scale $Q$. }

\appendix
\section{Matching onto Helicity Eigenstates}\label{helicityappendix}

We first review some standard features of the spinor helicity formalism (see, for example, \cite{Dixon:1996wi}).
We define spinors with momentum $p$ with definite helicity,
\begin{equation}
|p+\rangle\equiv p\rangle,\ |p-\rangle\equiv p], \langle p+|\equiv [p,\ \langle p-|\equiv \langle p.
\end{equation}
These have the explicit form\footnote{Note that the standard SCET convention of $p^\pm=p^0\mp p^3$ is the opposite of that used in \cite{Dixon:1996wi}.}
\begin{equation}\label{dirac}
\begin{aligned}
p\rangle = \frac{1}{\sqrt2}\left(\begin{matrix}\sqrt{p^-} \\ \sqrt{p^+}e^{i\phi_p}\\ \sqrt{p^-}\\ \sqrt{p^+}e^{i\phi_p}\end{matrix}\right) 
\end{aligned}
,\quad
\begin{aligned}
p] = \frac{1}{\sqrt2}\left(\begin{matrix}\sqrt{p^+}e^{-i\phi_p} \\ -\sqrt{p^-}\\ -\sqrt{p^+}e^{-i\phi_p}\\ \sqrt{p^-}\end{matrix}\right),
\end{aligned}
\end{equation}
where the phase is defined
\begin{equation}\begin{aligned}\label{phase}
e^{\pm i\phi_p} &= \frac{p^1\pm i p^2}{\sqrt{p^+p^-}}.
\end{aligned}\end{equation}
Inner products of spinors obey
\begin{equation}
\langle pq \rangle =[q p]^*, \langle p q\rangle=-\langle q p\rangle, \langle p q]=[pq\rangle =0.
\end{equation}
and
\begin{equation}
p\cdot q={1\over 2}\langle p q\rangle [q p].
\end{equation}
For positive energy spinors $p_i^0, p_j^0>0$, we have the explicit expressions
\begin{equation}\begin{aligned}
\langle p_i p_j\rangle&=\sqrt{p_i^+ p_j^-}e^{i\phi_i}-\sqrt{p_i^- p_j^+} e^{i \phi_j}\\
\left[p_i p_j\right]&=-\sqrt{p_i^+ p_j^-}e^{-i\phi_i}+\sqrt{p_i^- p_j^+} e^{-i \phi_j}
\end{aligned}\end{equation}
where the phases $\phi_j$ are defined in \eqn{phase}.
Helicity spinors obey the Schouten identity
\begin{equation}\label{schouten}
\langle p_i p_j\rangle \langle p_k p_l\rangle=\langle p_i p_k\rangle \langle p_j p_l\rangle+\langle p_i p_l\rangle \langle p_k p_j\rangle
\end{equation}
and Fierz rearrangement
\begin{equation}\label{fierz}
[p_i\gamma^\mu p_j\rangle [p_k\gamma^\mu p_l\rangle=2[p_i p_k]\langle p_l p_j\rangle.
\end{equation}
Charge conjugation gives 
\begin{equation}
[p\gamma^\mu q\rangle=\langle q \gamma^\mu p].
\end{equation} 
We also have, for any null vector $a^\mu$,
\begin{equation}\label{gammaspinor}
\fmslash{a}=a\rangle[a+a]\langle a .
\end{equation}
Finally, polarization vectors $\varepsilon^\mu$ are written in terms of spinors as
\begin{equation}\label{poldef}
\varepsilon^\mu_\pm(k,r)=\pm{\langle r\mp|\gamma^\mu|k\mp\rangle\over\sqrt{2}\langle r\mp|k\pm\rangle}
\end{equation}
where $r^\mu$ is an arbitrary lightlike reference momentum obeying $r\cdot p\neq 0$, corresponding to the gauge degree of freedom.  From (\ref{poldef}) and (\ref{fierz}) we then obtain
\begin{equation}
[p|\fmslash{\varepsilon}_-|q\rangle=-\sqrt{2}{[pr]\langle kq\rangle\over [rk]}
\end{equation}
and similiar expressions for the other helicities.

We can use the above results to simplify Feynman amplitudes between states of definite helicity.  It is convenient to introduce the spinors
$| n\pm\rangle$ and $| \bar n\pm\rangle$, defined with vanishing phase $\phi_{n,\nb}=0$.  The Schouten identity (\ref{schouten}) then allows us to express inner products of arbitrary spinors as inner products of spinors with $n\rangle$ and $\nb\rangle$:
\begin{equation}
\langle p q\rangle={\langle p\nb\rangle \langle qn\rangle-\langle p n\rangle\langle q\nb\rangle\over 2}
\end{equation}
where
\begin{equation}
\langle pn \rangle=\sqrt{2 p^+}e^{i\phi(p)},\ \langle p \nb\rangle=-\sqrt{2p^-}.
\end{equation}
To evaluate matrix elements of the form $\langle p\pm|\gamma^\mu|q\pm\rangle$, we decompose $\gamma^\mu$ via \eqn{gammadecomp}, and then from the definition \cite{Moult:2015aoa}
\begin{equation}
\xi^\mu_\pm={1\over 2\sqrt{2}} \langle n\pm|\gamma^\mu|\nb\pm\rangle
\end{equation}
we have
\begin{equation}
[p \fmslash{\xi}_+  q\rangle= {1\over2\sqrt{2}}[p\gamma^\mu q\rangle[n\gamma_\mu \bar n\rangle={1\over\sqrt{2}}[p n] \langle\nb q\rangle 
\end{equation}
and similarly for $\fmslash{\xi}_-$, which therefore gives
\begin{equation}\begin{aligned}
\ [p \gamma^\mu q\rangle={n^\mu\over 2}[p\nb]\langle \nb q\rangle+{\nb^\mu\over 2}[pn]\langle nq\rangle-{\xi_-^\mu\over\sqrt{2}}[pn]\langle\nb q\rangle-{\xi_+^\mu\over\sqrt{2}}[p\nb]\langle n q\rangle.
\end{aligned}\end{equation}
This immediately gives the zero-gluon matrix element 
\begin{equation}\begin{aligned}\label{fullQCDloapp}
{\cal M}_{q^\pm}
&\equiv  \langle p_2\pm | \gamma^\mu |p_1\pm\rangle\\
&=\pm\sqrt{2 p_1^- p_2^+} e^{\pm i\varphi_{p_1}}\eperp_{\mp}^{\mu}
\mp\sqrt{2p_1^+ p_2^-}e^{\mp i\varphi_{p_2}}\eperp_{\pm}^{\mu}
+\sqrt{p_1^+p_2^+}e^{\pm i(\varphi_{p_2}-\varphi_{p_1})}\;\nb^{\mu}+\sqrt{p_1^- p_2^-}\;n^{\mu}.
\end{aligned}\end{equation}

The one-gluon matrix elements between different helicity states can be read off from the corresponding Feynman diagrams and simplified using the above identities.  For example,
\begin{equation}\begin{aligned}
{\cal M}_{q^+g^+}&=-{g T^a}\left({[ p_2 \epslash^*_+(\pslash_2+\kslash)\gamma^\mu p_1\rangle\over 2\sp{p_2}{k}}-
{[p_2 \gamma^\mu(\pslash_1-\kslash)\epslash^*_+p_1\rangle\over 2\sp{p_1}{k}}\right)\\
&=-{g T^a}\left({[p_2 \epslash_- p_2\rangle [p_2\gamma^\mu p_1\rangle\over 2\sp{p_2}{k}}-
{[p_2 \gamma^\mu p_1\rangle [p_1\epslash_-p_1\rangle-[p_2 \gamma^\mu k\rangle [k\epslash_-p_1\rangle\over 2\sp{p_1}{k}}\right)\\
&={\sqrt{2}gT^a}\left(\left[{[p_2 r]\over[r k][p_2 k]}-{[p_1 r]\over[r k][p_1 k]}\right][p_2\gamma^\mu p_1\rangle-{1\over [p_1 k]}[p_2\gamma^\mu k\rangle\right)\\
&={\sqrt{2}gT^a}\left({[p_1 p_2]\over[p_2 k][p_1 k]}[p_2\gamma^\mu p_1\rangle-{1\over [p_1 k]}[p_2\gamma^\mu k\rangle\right)
\end{aligned}\end{equation}
where dependence on the gauge parameter $r$ has vanished. We can simplify this expression further by again using \eqref{gammadecomp} to decompose $\gamma^{\mu}$,
\begin{equation}\begin{aligned}\label{exact_result}
{\cal M}_{q^+g^+}&=\sqrt2 g T^a\Big(-\frac{\rsnp{p_2}{\nb} (\lsnp{\nb}{k} \rsnp{p_2}{k}+\lsnp{p_1}{\nb} \rsnp{p_2}{p_1})}{2 \rsnp{p_1}{k} \rsnp{p_2}{k}}n^{\mu}
-\frac{\rsnp{p_2}{n} (\lsnp{\nb}{k} \rsnp{p_2}{k}+\lsnp{p_1}{\nb} \rsnp{p_2}{p_1})}{\sqrt2 \rsnp{p_1}{k} \rsnp{p_2}{k}}\xi_{-}^{\mu}\\
&-\frac{\rsnp{p_2}{n} (\lsnp{n}{k} \rsnp{p_2}{k}+\lsnp{p_1}{n} \rsnp{p_2}{p_1})}{2 \rsnp{p_1}{k} \rsnp{p_2}{k}}\nb^{\mu}
-\frac{\rsnp{p_2}{\nb} (\lsnp{n}{k} \rsnp{p_2}{k}+\lsnp{p_1}{n} \rsnp{p_2}{p_1})}{\sqrt 2\rsnp{p_1}{k} \rsnp{p_2}{k}}\xi_{+}^{\mu}\Big)\\
&=\sqrt2 g T^a\Big(\frac{\rsnp{p_2}{n} \rsnp{p_2}{\nb} \sp{\nb}{q}}{2 \rsnp{p_1}{k} \rsnp{p_2}{k}}n^{\mu}
-\frac{\rsnp{p_2}{n} \rsnp{p_2}{\nb} \sp{n}{q}}{2 \rsnp{p_1}{k} \rsnp{p_2}{k}}\nb^{\mu}
+\frac{\rsnp{p_2}{n}{}^2 \sp{\nb}{q}}{\sqrt2 \rsnp{p_1}{k} \rsnp{p_2}{k}}\xi_{-}^{\mu}
-\frac{\rsnp{p_2}{\nb}{}^2 \sp{n}{q}}{\sqrt2 \rsnp{p_1}{k} \rsnp{p_2}{k}}\xi_{+}^{\mu}\Big),\\
\end{aligned}\end{equation}
where to get from the first to second line we've used a series of identities that can be derived from total momentum conservation. For example, since $p_1^{\mu}+q^{\mu} = p_2^{\mu}+k^{\mu}$, we have
\begin{equation}\label{totmom}
\fmslash{p}_1 + \frac{\sp{q}{n}}{2}\fmslash{\nb}+ \frac{\sp{q}{\nb}}{2}\fmslash{n} = \fmslash{p}_2+\fmslash{k},
\end{equation}
and we can sandwich \eqref{totmom} between arbitrary spinors, e.g. $[ p_2$ and $\nb \rangle$, to find
\begin{equation}
\lsnp{\nb}{k}\rsnp{p_2}{k}+\lsnp{p_1}{\nb}\rsnp{p_2}{p_1}=-\frac{\lsnp{n}{\nb}}{2}\sp{q}{\nb}\rsnp{p_2}{n}.
\end{equation}
Equation \eqref{exact_result} is now a particularly compact expression for the exact tree-level matrix element in QCD; however, it is useful to further manipulate it into a form that is simple to expand in the collinear limits of SCET. For example, let's set $p_1^{\perp}=0$, so that $k^{\perp}+p_2^{\perp} = 0$. This gives us
\begin{equation}
\sp{\xi_\pm}{k}+\sp{\xi_\pm}{p_2} = \frac{1}{2\sqrt2}\left(\lrsnpA{\nb}{k}\rlsnpA{n}{k} + \lrsnpA{\nb}{p_2}\rlsnpA{n}{p_2}\right) = 0,
\end{equation}
which, combined with the fact that $k^-+p_2^- = q^-$ when $p_1^-=0$, leads to the useful relation
\begin{equation}
\lrsnpA{p_2}{k} = \frac{\lrsnpA{p_2}{n}\lrsnpA{\nb}{k}-\lrsnpA{p_2}{\nb}\lrsnpA{n}{k}}{\lrsnpA{\nb}{n}}=\frac{2q\cdot \nb}{\lrsnpA{\nb}{n}}\frac{\lrsnpA{p_2}{n}}{\rlsnpA{k}{\nb}}.
\end{equation}
Also, on-shell conditions enforce that $k^+k^- = p_2^+p_2^-$ and $\lrsnpA{p_1}{\nb} = 0$, a concequence of the later being that
\begin{equation}
\lrsnpA{p_1}{k} = \frac{\lrsnpA{p_1}{n}\lrsnpA{\nb}{k}-\lrsnpA{p_1}{\nb}\lrsnpA{n}{k}}{\lrsnpA{\nb}{n}} = \frac{\lrsnpA{p_1}{n}\lrsnpA{\nb}{k}}{\lrsnpA{\nb}{n}}
\end{equation}
These identities allow us to further manipulate \eqref{exact_result} to 
\begin{equation}\begin{aligned}
{\cal M}_{q^+g^+}=&\sqrt{2}gT^a \Big(-\frac{\lsnp{\nb}{k} \rsnp{p_2}{\nb}}{\rsnp{\nb}{k} \rsnp{p_1}{n}}n^{\mu}+\frac{\lsnp{\nb}{k} \rsnp{p_2}{\nb} \sp{n}{q}}{\rsnp{\nb}{k} \rsnp{p_1}{n} \sp{\nb}{q}}\nb^{\mu}\\
&+\frac{\sqrt2 \lsnp{\nb}{k} \rsnp{p_2}{\nb}{}^2 \sp{n}{q}}{\rsnp{\nb}{k} \rsnp{p_1}{n} \rsnp{p_2}{n} \sp{\nb}{q}}\xi_{+}^{\mu}-\frac{\sqrt2 \lsnp{\nb}{k} \rsnp{p_2}{n}}{\rsnp{\nb}{k} \rsnp{p_1}{n}}\xi_{-}^{\mu}\Big).
\end{aligned}\end{equation}
In terms of boost-invariant variables, we finally have
\begin{equation}
{\cal M}_{q^+g^+}= -\sqrt{2}T^a g\frac{\sqrt{\sp{p_2}{\bar\eta}}}{\sqrt{\sp{p_1}{\eta}}}\Big(\bar{\eta }^{\mu}+\eta^{\mu}+\sqrt{2}e^{-i \phi (k)} \frac{\sqrt{\sp{p_2}{\eta}}}{\sqrt{\sp{p_2}{\bar{\eta }}}}\xi_{-}^{\mu}+\sqrt{2}e^{i \phi (k)}\frac{\sqrt{\sp{p_2}{\bar{\eta }}}}{\sqrt{\sp{p_2}{\eta}}}\xi_{+}^{\mu}\Big)
\end{equation}
Similar manipulations for the other helicities give the results in \eqn{fullqcdonegluon}.

\bibliography{bibliography}

\end{document}